\documentclass[12pt, a4paper]{article}

\pdfoutput=1

\usepackage{amsmath}
\usepackage{amsfonts}
\usepackage{amssymb}
\usepackage{bbm}
\usepackage{verbatim}
\usepackage{dsfont}
\usepackage{booktabs}
\usepackage{slashed}

\usepackage{epsfig}
\usepackage{color}
\usepackage[table]{xcolor}
\usepackage{graphicx}
\usepackage[]{caption}
\usepackage[listofformat=empty,subrefformat=empty]{subfig}
\usepackage{float}
\usepackage{overpic}

\usepackage{enumerate}
\usepackage{hhline}
\usepackage{multirow}

\usepackage{cite}
\usepackage{xspace}
\usepackage{setspace}

\usepackage[pdftitle={Chiral fermions and anomaly cancellation on orbifolds with Wilson lines and flux},
  pdfauthor={Wilfried Buchmuller, Markus Dierigl, Fabian Ruehle, Julian Schweizer},
  pdfsubject={orbifolds, Wilson lines, flux, anomalies},
  bookmarksopen, bookmarksnumbered, bookmarksopenlevel=2, colorlinks=false, linkcolor=blue, citecolor=blue, urlcolor=blue]{hyperref}

\DeclareMathOperator{\tr}{\text{tr}}

\newcommand{\comma}{\ensuremath{\enspace , \enspace}}

\renewcommand{\tfrac}{\genfrac{}{}{}1}

\begin{document}

\thispagestyle{empty}

\begin{flushright}
DESY-15-068\\
\end{flushright}
\vskip .8 cm
\begin{center}
{\Large {\bf Chiral fermions and anomaly cancellation\\[8pt] on orbifolds with Wilson lines and flux}}\\[10pt]

\bigskip
\bigskip 
{
{\bf{Wilfried Buchmuller}\footnote{E-mail: wilfried.buchmueller@desy.de}},
{\bf{Markus Dierigl}\footnote{E-mail: markus.dierigl@desy.de}},  
{\bf{Fabian Ruehle}\footnote{E-mail: fabian.ruehle@desy.de}},
{\bf{Julian~Schweizer}\footnote{E-mail: julian.schweizer@desy.de}}
\bigskip}\\[0pt]
\vspace{0.23cm}
{\it Deutsches Elektronen-Synchrotron DESY, 22607 Hamburg, Germany }\\[20pt] 
\bigskip
\end{center}

\begin{abstract}
\noindent
We consider six-dimensional supergravity compactified on orbifolds with Wilson lines and bulk flux. Torus Wilson lines are decomposed into Wilson lines around the orbifold fixed points, and twisted boundary conditions of matter fields are related to fractional localized flux. Both, orbifold singularities and flux lead to chiral fermions in four dimensions. We show that in addition to the standard bulk and fixed point anomalies the Green-Schwarz term also cancels the four-dimensional anomaly induced by the flux background. The two axions contained in the antisymmetric tensor field both contribute to the cancellation of the four-dimensional anomaly and the generation of a vector boson mass via the Stueckelberg mechanism. An orthogonal linear combination of the axions remains massless and couples to the gauge field in the standard way. Furthermore, we construct convenient expressions for the wave functions of the zero modes and relate their multiplicity and behavior at the fixed points to the bulk flux quanta and the Wilson lines.
\end{abstract}

\newpage 
\setcounter{page}{2}
\setcounter{footnote}{0}
\tableofcontents

{\renewcommand{\baselinestretch}{1.3}

\section{Introduction}
\label{sec:Introduction} 

It is a well known puzzle of the Standard Model of particle physics that quarks and leptons come in three copies of complete representations of the grand unified (GUT) group $SU(5)$, whereas the Higgs doublet and the gauge fields form incomplete or ``split'' GUT representations. This raises the question why the spectrum of light scalars and gauge fields in the Standard Model directly
reflects the breaking of the GUT symmetry whereas, on the other hand, the light fermions still show the unbroken underlying GUT symmetry, accompanied by a threefold replication.

Ingredients of a possible answer to this question can be found in higher-dimensional supersymmetric gauge and string theories\footnote{For detailed discussions and extensive references see, for example, \cite{GreenSchwarzWitten,Ibanez:2012zz}.}. These theories can lead to a chiral spectrum of fermions due to background gauge fields or singularities, as in orbifold theories. The corresponding anomalies can be canceled by the Green-Schwarz mechanism \cite{Green:1984sg}. The presence of flux naturally leads to a multiplicity of fermion zero modes \cite{Witten:1984dg}, and Wilson lines allow for various patterns of GUT symmetry breaking which can be associated with the appearance of split multiplets \cite{Witten:1985xc}. 

In this paper we study some aspects of supergravity in six dimensions \cite{Nishino:1984gk,Nishino:1986dc} compactified on $T^2$ and one of its orbifolds, $T^2/\mathbb{Z}_2$. Orbifold field theories in five and six dimensions have already been successfully used to construct models of grand unification (see e.g.~\cite{Kawamura:2000ev,Hall:2001pg,Hebecker:2001wq,Asaka:2001eh}), and GUT scale extra dimensions can indeed arise as an intermediate step in anisotropic compactifications of string theories \cite{Hebecker:2004ce,Buchmuller:2007qf}. However, in all of these models possible effects of bulk flux were ignored.

In six-dimensional field theories gauge flux plays an important role for supersymmetry breaking \cite{Bachas:1995ik} and moduli stabilization \cite{Braun:2006se}. In addition, flux leads to a chiral spectrum of fermions in four dimensions. It is then an important question of theoretical consistency how the corresponding anomalies, which are not included in the familiar bulk and fixed point anomalies, are canceled. In the following, we study in detail how this is achieved by means of the Green-Schwarz mechanism. An important related question is the generation of a mass for the anomalous $U(1)$ vector boson to which now two axions contribute. Contrary to the standard Stueckelberg mechanism \cite{Stueckelberg:1900zz}, the vector boson mass is generated by both, an axion-vector boson coupling contained in the Green-Schwarz term and the classical flux. We also reconsider the connection between twisted boundary conditions and discrete Wilson lines. A simple picture for the possible patterns of boundary conditions is obtained in terms of closed Wilson lines around the orbifold fixed points. Finally, we evaluate convenient expressions for the wave functions of charged matter fields \cite{Cremades:2004wa,Abe:2013bca} which are relevant for the calculation of Casimir energies on orbifolds \cite{Ponton:2001hq,Ghilencea:2005vm,Braun:2006se,Buchmuller:2009er}.

The paper is organized as follows. Some aspects of the orbifold geometry are reviewed in Sec.~\ref{sec:InternalSpace}. Wilson lines on a torus and an orbifold, with and without flux, and their connection with twisted boundary conditions are discussed in Sec.~\ref{sec:WilsonLines}. In Secs.~\ref{sec:6dSugra} and \ref{sec:Anomalies} we evaluate the effective action of moduli and gauge fields, and we study the Green-Schwarz mechanism and the realization of the Stueckelberg mechanism. The wave functions for fermion zero-modes are investigated in Sec.~\ref{sec:Wavefunctions}. Summary and outlook are given in Sec.~\ref{sec:Conclusion}.

\section{The internal space}
\label{sec:InternalSpace}

We consider supergravity theories in six dimensions, two of which are compactified. As background geometry we assume the product $M \times X$ of Minkowski space and some internal space $X$ with the metric
\begin{align}
(g_6)_{MN} = \begin{pmatrix} r^{-2} (g_4)_{\mu \nu} & 0 \\ 0 & r^2
  (g_2)_{mn} \end{pmatrix} \,,
\label{6dmetric}
\end{align}
where $\mu,\nu = 0 \dots 3$ and $m,n = 5,6$. Instead of the coordinates $(x^5,x^6)$ of the internal space we mainly use rescaled, dimensionless coordinates $y=(y_1,y_2) = (x^5, x^6)/L$, where $L$ denotes a fixed, physical length scale. Indices are raised and lowered with the metrics $g_6$, $g_4$ and $g_2$, respectively. The additional rescaling by the dimensionless radion field $r$ leads to standard kinetic terms for the moduli and parametrizes the size of the extra dimensions.

In the following, we consider the two-dimensional torus $T^2$ and one of its orbifolds, $T^2/\mathbb{Z}_2$. To set our notations and conventions and to connect the properties of the two internal spaces we briefly review their geometry with emphasis on a convenient basis of 1-cycles.

The two-dimensional torus $T^2$ is obtained by starting from the universal covering space $\mathbb{R}^2$ and modding out a two-dimensional lattice. This lattice is generated by two vectors $\lambda_{1,2}$ corresponding to the basic translations $t_{1,2}$. They describe the fundamental domain. An arbitrary lattice vector $\lambda$ can be parametrized as a linear combination with integer coefficients, $\lambda = n_1 \lambda_1 + n_2 \lambda_2$. This gives rise to an equivalence relation which can be expressed in terms of the torus coordinates $y = (y_1,y_2)$ as
\begin{align}
y \sim y + \lambda \,,
\end{align}
and induces a $\mathbb{Z}^2$ group of translations. The corresponding 1-cycles and the fundamental domain of the torus are depicted in Fig.~\ref{toruslattice}. In these coordinates, the shape of the torus is encoded via two real shape moduli $\tau_{1,2}$ in the two-dimensional metric $(g_2)_{mn}$,
\begin{align}
(g_2)_{mn} = \frac{1}{\tau_2} \begin{pmatrix} 1 & \tau_1 \\ \tau_1 & \tau_1^2 + \tau_2^2 \end{pmatrix} \,.
\end{align}
The shape moduli can be combined into the complex parameter $\tau = \tau_1 + i \tau_2$ and its complex conjugate $\bar{\tau}$. The physical volume of the torus is $V_{T^2} = r^2 L^2$.

\begin{figure}
\centering
\begin{overpic}[scale = 0.4, tics=10]{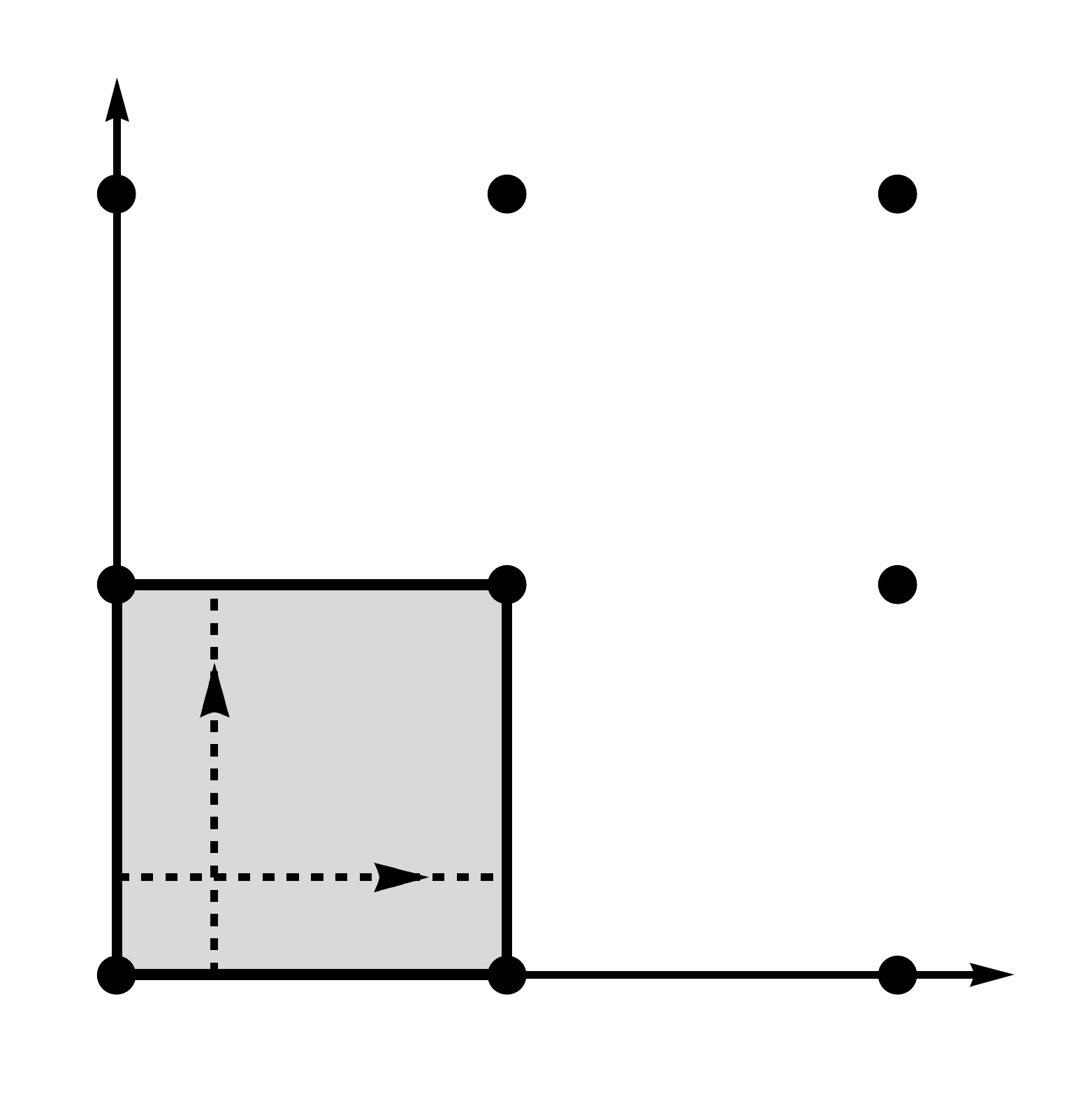}
	\put(95, 10){$y_1$}
	\put(8, 95){$y_2$}
	\put(33, 23){$\mathcal{T}_1$}
	\put(22, 35){$\mathcal{T}_2$}
	\put(5, 3){$0$}
	\put(45, 3){$1$}
	\put(81, 3){$2$}
	\put(5, 45){$1$}
	\put(5, 81){$2$}
\end{overpic}
\caption{A torus lattice $\mathbb{Z}^2$ and its fundamental domain (gray); the canonical basis of 1-cycles $\mathcal{T}_{1,2}$ is depicted by dashed arrows.}
\label{toruslattice}
\end{figure}
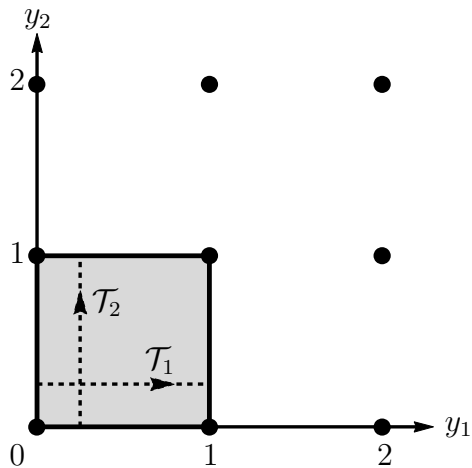

For the orbifold $T^2/\mathbb{Z}_2$ we have to further mod out a $\mathbb{Z}_2$ rotational symmetry generated by $p$, which acts on the internal space via
\begin{align}
p:~ y \rightarrow -y \comma p^2 = 1 \,.
\end{align}
As a consequence, the fundamental domain of the orbifold has half the area of the torus and is completely covered by $y_1 \in [0, 1/2]$ and $y_2 \in [0, 1)$. Thus, its volume is $V_{T^2/\mathbb{Z}_2} = \tfrac{1}{2} r^2 L^2$. The extended group acting on the internal coordinates, the so-called space group, is generated by $\{t_1, t_2, p\}$. It is the semi-direct product $\mathbb{Z}^2 \rtimes \mathbb{Z}_2$, whose structure becomes apparent in the relation 
\begin{align}
p \, t_m p = (t_m)^{-1} \,.
\label{parityrelation}
\end{align}
This follows from considering an element of the space group, $p \, t_m p\, t_m$, which acts as the identity on the universal covering space. 

The space group does not act freely, i.e.\ there are fixed points located at
\begin{align}
\zeta_1 = (0, 0) \comma \zeta_2 = (1/2, 0) \comma \zeta_3 = (0, 1/2) \comma \zeta_4 = (1/2, 1/2) \,.
\end{align}
The orbifold thus has the topology of a sphere with four points removed. At each fixed point there is a conical singularity with deficit angle $\pi$. This can be interpreted as a singular curvature on the fixed points. The region away from the fixed points, the bulk, is flat.

We can now study the orbifold 1-cycles and their decomposition in terms of the torus 1-cycles. Orbifold 1-cycles wind around the fixed points. The canonical basis is given by three 1-cycles $\mathcal{C}_{1,2,3}$ encircling the associated fixed point once, cf.\ Fig.~\ref{fundamentaldomain}. The $\mathbb{Z}_2$ operator $p$ corresponds to the 1-cycle $\mathcal{C}_1$. Note that the 1-cycle around the fourth fixed point $\mathcal{C}_4$ is not linearly independent,
\begin{align}
\mathcal{C}_4 = -( \mathcal{C}_1 + \mathcal{C}_2 + \mathcal{C}_3) \,,
\end{align}
where the minus sign signals a reversed orientation.

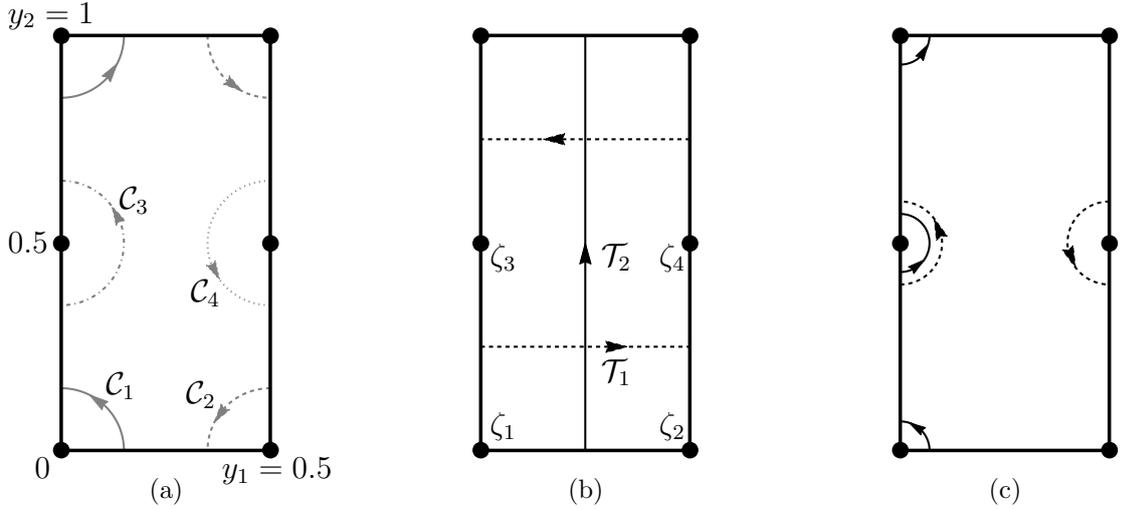
\begin{figure}
\centering
	\subfloat[\label{fundamentaldomain}]{
		\begin{overpic}[width= 0.2\textwidth]{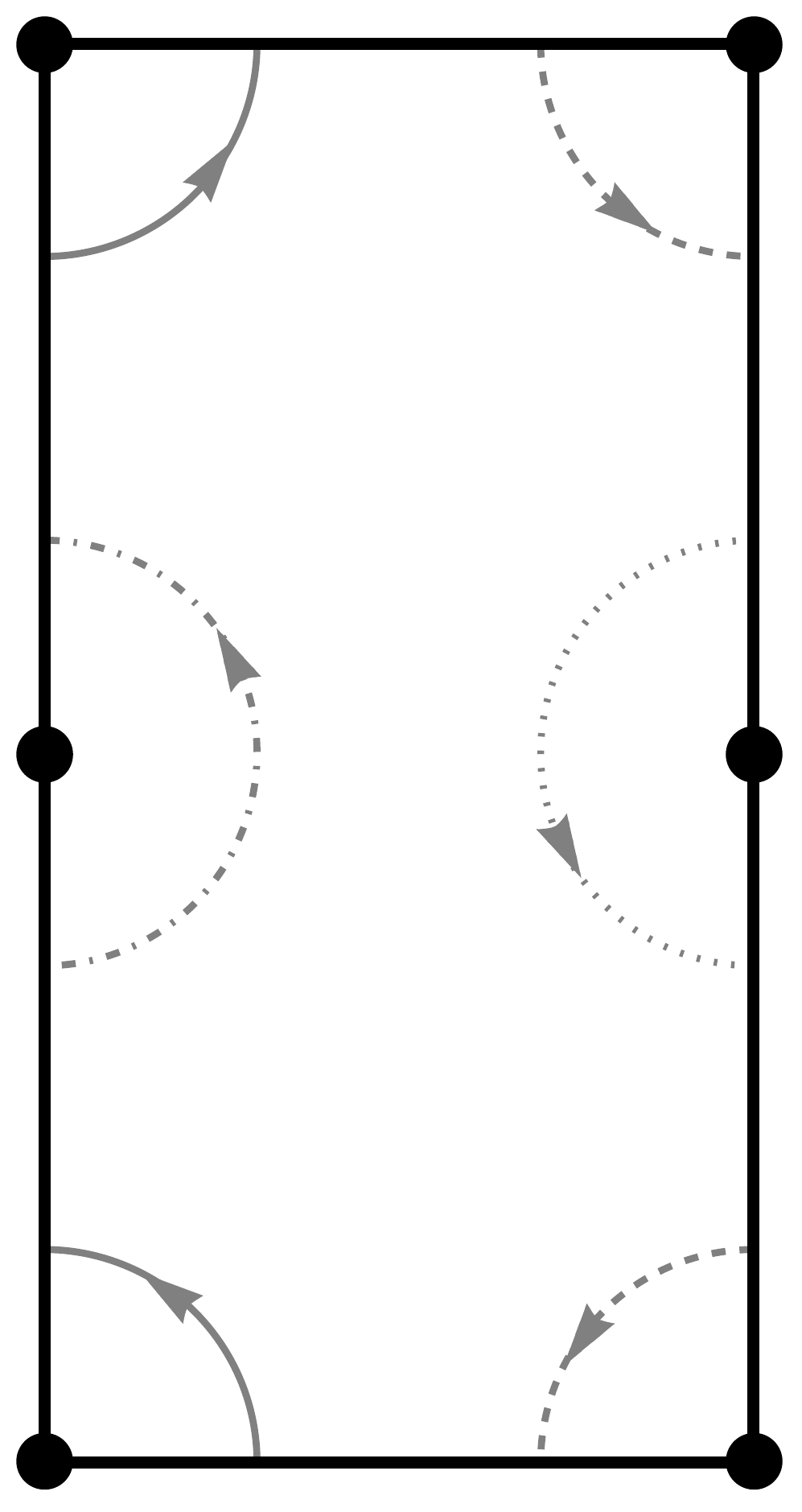}
			\put(13,16){$\mathcal{C}_1$}
			\put(31,14){$\mathcal{C}_2$}
			\put(16,58){$\mathcal{C}_3$}
			\put(32,37){$\mathcal{C}_4$}
			\put(-3,-3){$0$}
			\put(-9,48){$0.5$}
			\put(-9,100){$y_2 = 1$}
			\put(39,-3){$y_1 = 0.5$}
		\end{overpic}
	}
	\hspace{2cm}
	\subfloat[\label{projection}]{
		\begin{overpic}[width=0.2 \textwidth]{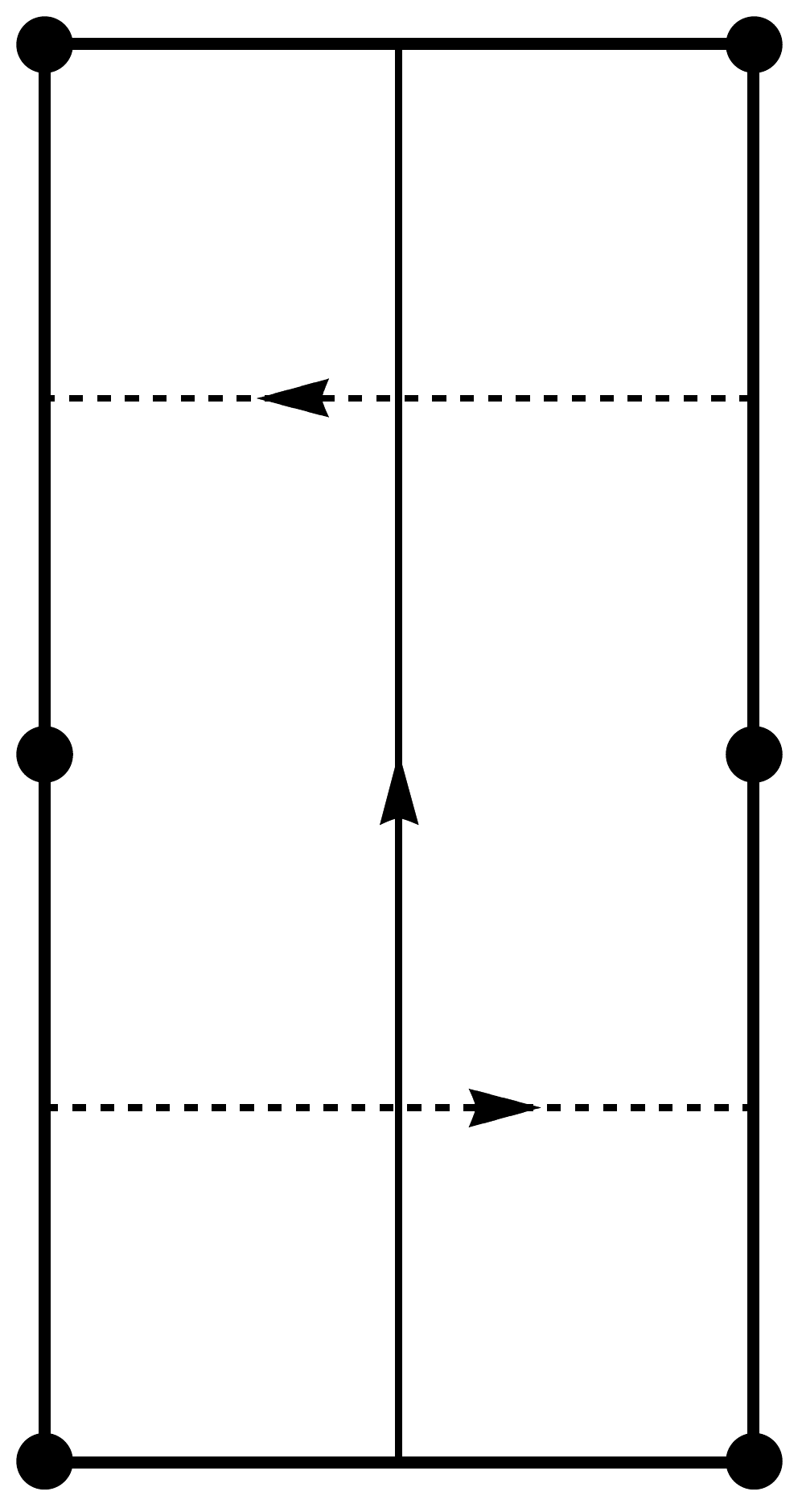}
			\put(5,7){$\zeta_1$}
			\put(43,7){$\zeta_2$}
			\put(5,45){$\zeta_3$}
			\put(43,45){$\zeta_4$}
			\put(30,45){$\mathcal{T}_2$}
			\put(30,19){$\mathcal{T}_1$}
	\end{overpic}
	}
	\hspace{2cm}
	\subfloat[\label{projectiondeformation}]{
		\begin{overpic}[width= 0.2\textwidth]{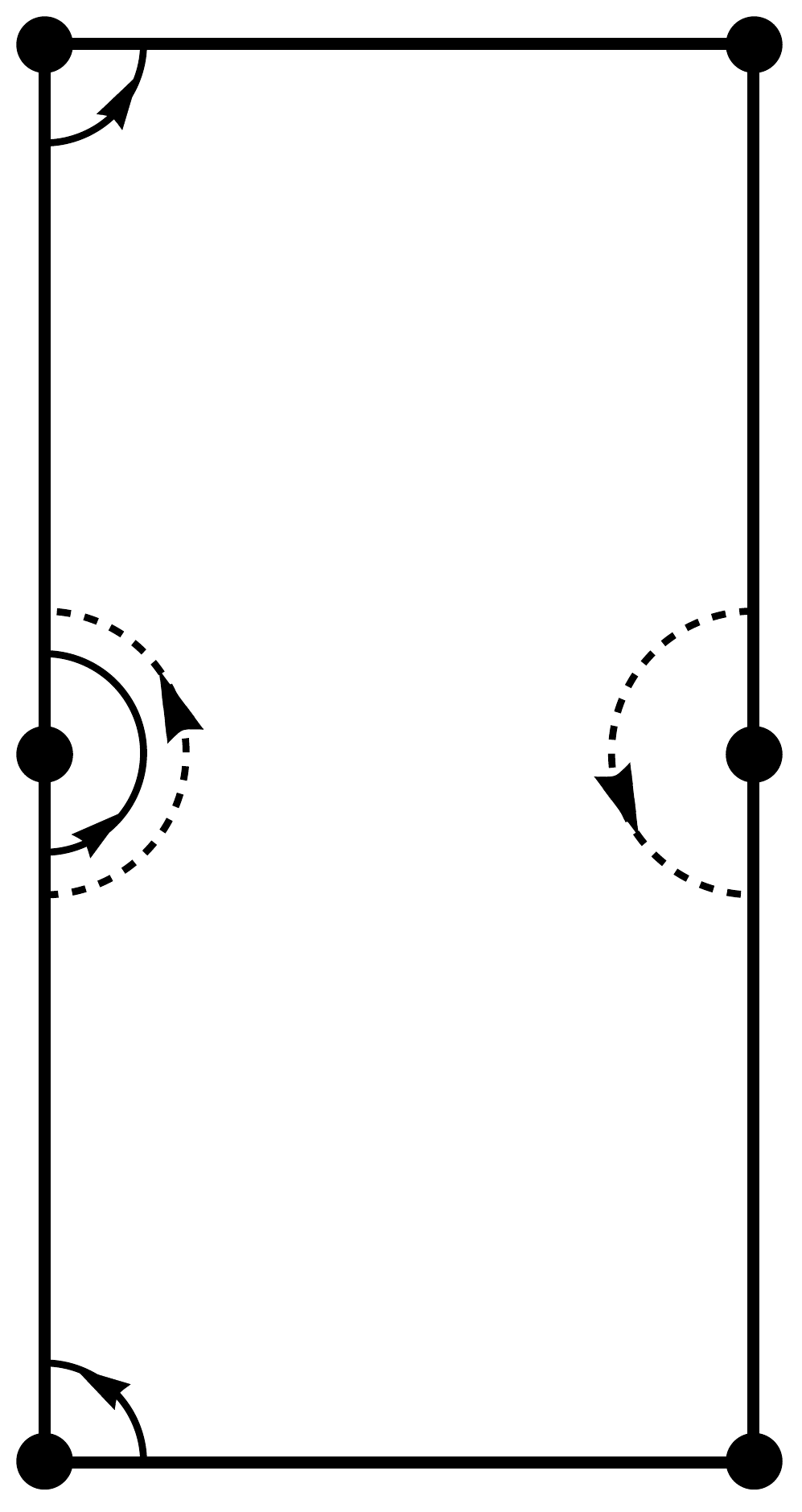}
		\end{overpic}
	}
	\caption{The fundamental domain of $T^2/\mathbb{Z}_2$; the
          black dots denote the orbifold fixed points.  The canonical
          basis of the orbifold 1-cycles is presented in
          (a). (b) shows the projection of the torus 1-cycles $\mathcal{T}_{1,2}$, and (c) illustrates the decompositions
$\mathcal{T}_1 = \mathcal{C}_3+ \mathcal{C}_4$ and $\mathcal{T}_2 = \mathcal{C}_1+ \mathcal{C}_3$.}
	\label{cyclerelation}
\end{figure}

In order to express the torus cycles in terms of the canonical basis we project them into the fundamental domain of the orbifold, see Fig.~\ref{projection}, and then deform them continuously, see Fig.~\ref{projectiondeformation}. We thus obtain the decomposition
\begin{align}
\mathcal{T}_1 = \mathcal{C}_3 + \mathcal{C}_4 = - (\mathcal{C}_1 + \mathcal{C}_2) \,,\quad \mathcal{T}_2 = \mathcal{C}_1 + \mathcal{C}_3 \,.
\end{align}
The geometry described above has physical consequences for the fields of the theory, which are discussed in the next section.

\section{Wilson lines}
\label{sec:WilsonLines}

Given the background geometry of the internal space, we can study its effect on the quantum fields of the theory. On non-simply connected spaces such as $T^2$ or $T^2/\mathbb{Z}_2$ there is the possibility to introduce twisted periodicity conditions with respect to the global symmetry group $\mathcal{G}$ of the fields. If the global symmetry is part of a gauge symmetry, the twisted periodicity conditions can be related to background configurations of the gauge fields, so-called Wilson lines. In the following, we restrict our discussion to a $U(1)$ gauge theory, distinguishing the cases with and without bulk flux.

\subsection{Twists and Wilson lines}
\label{subsec:GlobalTwists}

We start by reviewing the basic concepts for twists and Wilson lines on $T^2$ and $T^2/\mathbb{Z}_2$. Let $\phi$ be an arbitrary field of the theory defined on the universal covering space. The projection to the compact internal space leads to the identification of field values at different points in $\mathbb{R}^2$ related by the action of a space group element $s$ \cite{Witten:1985xc,Hebecker:2001jb}, 
\begin{align}
\phi\left(s(y)\right) = \phi(y) \,.
\end{align}
This consistency condition leads, among other things, to the usual Kaluza-Klein spectrum in the compactified theory. In general, the field $\phi$ transforms under a group of global symmetries $\mathcal{G}$. These may contain global parts of gauge symmetries as well as R- or flavor symmetries. As a consequence, the periodicity conditions can be augmented by so-called twists $T_{s} \in \mathcal{G}$ such that
\begin{align}
\phi\left(s(y)\right) = T_{s} \, \phi(y) \,.
\label{globaltwist}
\end{align}
The twists $T_{s}$ may further depend on the representation of the fields with respect to the global symmetry group.\footnote{For a $U(1)$ group this corresponds to the charge $q$.} The fields have
to be well-defined and single-valued on the universal covering space, which means that the map from the space group into $\mathcal{G}$ has to be a group homomorphism \cite{Hebecker:2001jb}.

For the orbifold there is an additional generator $p$ of the space group, which corresponds to the element $T_p \equiv P \in \mathcal{G}$. Using Eq.~\eqref{parityrelation} together with the Abelian nature of the gauge group, one has
\begin{align}
(T_m)^{-1} = P T_m P = P^2 T_m  = T_m \quad \text{i.e.,} \quad (T_m)^2 = 1 \,.
\label{abeliantwists}
\end{align}
This leads to a severe restriction for the allowed twists under translations by lattice vectors. Below we discuss this for a $U(1)$ gauge symmetry and relate it to continuous and discrete Wilson lines on $T^2$ and $T^2/\mathbb{Z}_2$, respectively.

As mentioned above, the values of fields at points in $\mathbb{R}^2$ related by a space group action have to be identified. This leads to equivalence classes of field values on the internal space. Two values of the field $\phi$ and $\tilde{\phi}$ are equivalent if there is an element $s$ in the space group such that
\begin{align}
T_s \, \phi = \tilde{\phi} \,.
\end{align}
Thus, not only the internal space gets modified by the space group action but also the range of the fields fulfilling twisted boundary conditions.

Let us briefly comment on the influence of different representations under the Lorentz group of the internal space. On the torus this does not change the above considerations due to its flatness. On the orbifold, however, the fixed points carry singular curvature. Consequently, different Lorentz representations transform differently under a parallel transport around the fixed points. It is a subtle issue to arrange the global twists in order to obtain a consistent field spectrum in the four-dimensional effective theory. Concretely, one has to involve different R-symmetry twists in order to retain fermion zero-modes in gauge- and hypermultiplets, see e.g.\ \cite{Lee:2005tk,Braun:2006se}. This is essential in order to preserve part of the supersymmetry. We will not go into further detail here and only consider the overall signs created by twists in the global symmetry group that lead to chiral boundary conditions introduced in Sec.~\ref{sec:Anomalies}. The additional effect of Wilson lines is treated explicitly.

For an Abelian gauge theory on the torus the commutativity condition of the twist operators $T_{1,2}$ is trivially satisfied. Thus, one can choose two arbitrary $U(1)$ elements parametrized by real parameters $\alpha_m$,
\begin{align}
T_m = e^{iq \alpha_m} \,.
\label{twistoperator}
\end{align}
Here, $q$ denotes the charge of the twisted field under the $U(1)$ gauge group. The inequivalent choices for $\alpha_m$ take values in the interval
\begin{align}
\alpha_m \in [0, 2\pi/q) \,.
\label{independenttwists}
\end{align}
Further, they are subject to the identification $\alpha_m \sim \alpha_m + 2\pi/q$. For theories with fields of different charges the classification of independent twists follows from \eqref{independenttwists} by using the smallest charge in the spectrum.

The action of the twist operators on a field $\phi$ with charge $q$, defined on the universal covering space $\mathbb{R}^2$, reads\footnote{On $T^2$ the allowed transformations are translations by lattice vectors $\lambda = n \lambda_1 + m \lambda_2$.}
\begin{align}
\phi\left(y + \lambda \right) = T_{\lambda} \, \phi(y) = e^{iq (n \alpha_1 + m \alpha_2)} \phi(y) \,.
\label{periodicitymod}
\end{align}
Under gauge transformations the charged fields transform as
\begin{align}
\phi(y) \rightarrow e^{iq\Lambda(y)} \phi(y) \,.
\end{align}
The local gauge parameter $\Lambda(y)$ is an arbitrary, real function on $\mathbb{R}^2$. The gauge field $A$ transforms as usual, $A \rightarrow A - d \Lambda$. Its associated gauge invariant field strength is denoted by $F = dA$. Using the dimensionless coordinates $y_m$, the gauge field components $A_m$ are dimensionless, as well. To restore their physical mass dimension they have to be multiplied by $1/L$; note that the 1-form $A$ is not affected. Similar reasoning holds for the field strength $F$.

We first consider the case where $F=0$. Performing a gauge transformation on the universal covering space with a gauge parameter obeying
\begin{align}
\Lambda(y + \lambda) = - (n \alpha_1 + m \alpha_2) \Lambda(y) \,,
\label{periodgauge}
\end{align}
leads to trivial periodicity conditions for the gauge transformed fields
\begin{align}
e^{iq \Lambda(y + \lambda)} \phi(y + \lambda) = e^{iq \Lambda(y)} \phi(y) \,.
\end{align}
The gauge field, on the other hand, develops a background value. With the convenient choice $\Lambda(y) = -(\alpha_1 y_1 + \alpha_2 y_2)$, satisfying Eq.~\eqref{periodgauge}, it reads
\begin{align}
A = -d\Lambda = \alpha_m dy_m \,.
\label{backgroundgauge}
\end{align}
The associated Wilson lines are described in the usual way, by the path-ordered exponential of the gauge field along a (possibly open) path $\mathcal{T} \subset \mathbb{R}^2$,
\begin{align}
W = \mathcal{P} \exp{\left[ i q \int_{\mathcal{T}} A \right]} \in U(1) \,.
\end{align}
If the path connects points separated by a lattice vector $\lambda$, the Wilson line transforms under the gauge transformations Eq.~\eqref{periodgauge} as
\begin{align}
W \rightarrow e^{iq \Lambda(y)} W e^{- iq \Lambda(y+\lambda)} = W e^{iq(n \alpha_1 + m \alpha_2)}\,,
\label{wilsontrafo}
\end{align}
where $y$ denotes the starting point of $\mathcal{T}$. Therefore, on the universal covering space one can relate twisted boundary conditions for charged fields and constant background gauge configurations. This has physical consequences once one projects to the compact internal space.

\subsection{Wilson lines and flux}
\label{subsec:WilsonWithout}

The constant gauge field \eqref{backgroundgauge} satisfies the periodicity condition on the torus. As a result it can be straightforwardly projected. But even though it was a pure gauge configuration on $\mathbb{R}^2$ and hence unobservable, it is physically relevant on $T^2$. This is due to the fact that valid gauge transformations on the internal space are restricted to fulfill periodicity conditions.

In order to be well-defined the phases for charged fields generated by gauge transformations have to coincide on points related by a space group action. On $T^2$ this means
\begin{align}
e^{iq \Lambda(y + \lambda)} = e^{iq \Lambda(y)} \,.
\end{align}
Accordingly, the local gauge parameter $\Lambda$ is constrained by
\begin{align}
\Lambda(y + \lambda) = \Lambda(y) + \frac{2\pi k}{q} \comma k \in \mathbb{Z} \,.
\end{align}
For $k \neq 0$ these are the large gauge transformations. Without loss of generality we can parametrize the relevant transformations on $T^2$ by two integers $k_{1,2}$,
\begin{align}
\Lambda(y) = \frac{2\pi}{q} (k_1 y_1 + k_2 y_2) \,.
\end{align}
These gauge transformations correspond to the gauge field backgrounds
\begin{align}
A = \left(\alpha_m + \frac{2\pi k_m}{q} \right) dy_m\,.
\end{align}
This is directly related to the inequivalent twists in Eq.~\eqref{twistoperator}. Furthermore, the restriction of gauge transformations on the torus ensures gauge invariance of the Wilson lines around 1-cycles, see~\eqref{wilsontrafo}. The Wilson lines, or equivalently the gauge twists on $T^2$, are described by two continuous parameters $\alpha_m$ in the range of \eqref{independenttwists}. They describe physically different background configurations, the continuous Wilson lines on the torus. As long as $F=0$, the closed path $\mathcal{T}$ around which the Wilson loop is evaluated can be continuously deformed. 

In the following we discuss the framework of canonical Wilson lines on orbifolds and illustrate its advantages in the presence of bulk flux. In particular, we explicitly construct the gauge field configuration in terms of vortex solutions. Modding out the rotational symmetry restricts the allowed values of the Wilson lines, see Eq.~\eqref{abeliantwists}. The possible twist operators $T_m$, $P$ are encoded in terms of phases, in analogy to Eq.~\eqref{twistoperator} 
\begin{align}
\alpha_m = \frac{k_m \pi}{q}\,,\quad\alpha_p = \frac{k_p \pi}{q}\,,\qquad k_m,k_p \in \mathbb{Z} \,.
\label{translationphase}
\end{align}
The parameters $k_m$ and $k_p$ are integers and hence the orbifold projection restricts the possible twists to a discrete set. Our aim is to express these twists as a background configuration of the gauge field that is well-defined on $T^2/\mathbb{Z}_2$. The value of the twist operators $T_1, T_2, P$ are then obtained by evaluating the Wilson lines for the corresponding 1-cycles $\mathcal{T}_1, \mathcal{T}_2, \mathcal{C}_1$.

In order to obtain a well-defined theory for the charged fields, the covariant derivative $D_m = \partial_m + iq A_m$ has to transform as the ordinary derivative, leading to a sign change of the internal components of the gauge field under the $\mathbb{Z}_2$ transformation
\begin{align}
P:~ A_m \rightarrow - A_m \,.
\label{gaugeparity}
\end{align}
Hence, the constant gauge background~\eqref{backgroundgauge} is not allowed on the orbifold. In fact, an appropriate gauge field should not create any flux in the bulk of the orbifold, generate the desired phases around the (projected) 1-cycles $\mathcal{T}_1$, $\mathcal{T}_2$, $\mathcal{C}_1$, and satisfy condition \eqref{gaugeparity} in the bulk. The correct configuration is a superposition of vortex solutions in the universal covering space that are centered around the positions of the fixed points. In the vicinity of the vortex center $\zeta$ it has the form \cite{Abrikosov:1956sx, Nielsen:1973cs}
\begin{align}
A_m^{\text{vort}}(y; \zeta,c) = - \frac{c}{|y - \zeta|^2} \epsilon_{mn} (y_n - \zeta_n) \,,
\end{align}
where $c$ is a real constant that parametrizes the phase generated for charged fields around the vortex, as in the Aharonov-Bohm effect. It plays a role analogous to that of $\alpha_m$ on the torus. On the covering space the complete solution is a sum of vortices around the fixed point positions
\begin{align}
A_m^{\text{orb}}(y; c_i) = \sum_{\lambda \in \mathbb{Z}^2} \left( \sum_{i = 1}^4 A^{\text{vort}}_m (y; \zeta_i + \lambda, c_i) \right) \,.
\label{orbbackground}
\end{align}
Away from the fixed points the field strength $F$ vanishes, as desired. 

Moreover, the correct transformation behavior of $A_m^{\text{orb}}$ under $P$ is ensured. Since
\begin{align}
P:~ A_m^{\text{vort}} (y; c, \zeta) \rightarrow  -A_m^{\text{vort}}(y; c, -\zeta) \,,
\end{align}
we find from Eq.~\eqref{orbbackground} that
\begin{align}
P:~ A_m^{\text{orb}}(y; c_i) \rightarrow -A_m^{\text{orb}}(y; c_i) \,,
\end{align}
where we have used that $2 \zeta_i$ is a lattice vector. Therefore, $A^{\text{orb}}$ can be projected onto the orbifold. This gauge field configuration allows to express the phases of charged fields as Wilson lines around the orbifold 1-cycles. The relevant phases read
\begin{align}
\oint_{\mathcal{C}_i} A^{\text{orb}}(y; c_i) = \pi c_i \,.
\end{align}
For the Wilson lines corresponding to the basis of orbifold 1-cycles this yields
\begin{align}
W_i = \exp{\left( i q \oint_{\mathcal{C}_i} A^{\text{orb}} \right)} = e^{i q \pi c_i  } \,,
\label{sec:wilsonlines:eq:Wilson_line_phase}
\end{align}
and in order to satisfy Eq.~\eqref{translationphase} we find
\begin{align}
c_i = \frac{k_i}{q} \,,
\end{align}
with integer parameters $k_i \in \mathbb{Z}$. Hence, the Wilson lines take the values $\pm 1$ and only a discrete set of valid gauge field configurations exists. Thus, Wilson lines on the orbifold are discrete.

Since $\mathcal{C}_4$ is a linear combination of the basis of 1-cycles, we find an additional consistency condition
\begin{align}
\pi c_4 = \oint_{\mathcal{C}_4} A^{\text{orb}} = - \sum_{i=1}^3 \oint_{\mathcal{C}_i} A^{\text{orb}} = - \pi \sum_{i=1}^3 c_i \,,
\label{consistencygauge}
\end{align}
which, on the level of Wilson lines, yields
\begin{align}
W_4 = \prod_{i=1}^{3} W_i \,.
\end{align}
Hence, instead of describing periodicity conditions for charged fields on orbifolds via the twist operators $P$, $T_1$, and $T_2$, we can use a more convenient description in terms of Wilson lines $W_i$ around the orbifold 1-cycles. We list the relation between the two conventions for the eight possible configurations on $T^2/\mathbb{Z}_2$ in Tab.~\ref{Tab:Phases}. The classification in terms of Wilson lines around the fixed points generalizes to other $T^2/\mathbb{Z}_N$ orbifolds. There the advantages become even more apparent, since the description in terms of torus Wilson lines is redundant whereas the orbifold 1-cycles are not.

\begingroup
\small
\begin{table}
\centering
\begin{tabular}[c]{|c||c|c|c|c|}
\hline
$(P, T_1, T_2)$ & $(+,+,+)$ & $(+,+,-)$ & $(+,-,+)$ & $(+,-,-)$\\
\hline
$(W_1,W_2,W_3)$ & $(+,+,+)$ & $(+,+,-)$ & $(+,-,+)$ & $(+,-,-)$\\
\hline
\end{tabular}\\[2mm]
\noindent
\begin{tabular}[c]{|c||c|c|c|c|}
\hline
$(P, T_1, T_2)$ & $(-,+,+)$ & $(-,+,-)$ & $(-,-,+)$ & $(-,-,-)$\\
\hline
$(W_1,W_2,W_3)$ & $(-,-,-)$ & $(-,-,+)$ & $(-,+,-)$ & $(-,+,+)$\\
\hline
\end{tabular}
\caption{Relation between the phases $\{P, T_1, T_2\}$ and $\{W_1, W_2, W_3\}$ generated by discrete Wilson lines.}
\label{Tab:Phases}
\end{table}
\endgroup

The gauge field background~\eqref{orbbackground} and the associated phases further suggest an alternative interpretation in terms of localized, singular gauge fluxes at the fixed points, see also \cite{vonGersdorff:2006nt}. $W_i = -1$ indicates a flux located at $\zeta_i$ that generates a phase factor $(-1)$ for fields with electric charge $q$. For a general configuration described above, this yields
\begin{align}
F = \sum_{i=1}^4 \left[ \frac{2\pi}{q}\left( \frac{1}{2} \, \delta_{(W_i, -1)} + k_i \right) \delta(y- \zeta_i) \right] v_2 \,,  \quad k_i \in \mathbb{Z}\,,
\label{wilsonflux}
\end{align}
where $v_2=dy_1 \wedge dy_2$ denotes the volume element and $\delta_{(W_i, -1)}$ is the Kronecker delta. This formulation is especially useful for the discussion of fixed point anomalies in a Wilson line background and orbifolds $T^2/\mathbb{Z}_N$ with $N > 2$, that will be discussed in future work \cite{bdrs1}. Note that, while the presence of localized fluxes is obscured using torus cycles, they arise naturally in the above description. The consistency condition~\eqref{consistencygauge} applied to the flux reads
\begin{align}
\sum_{i=1}^4 \left( \frac{1}{2} \delta_{(W_i, -1)} + k_i \right) = 0 \,.
\end{align}
Note that while the flux on the orbifold fixed points is fractional the vortex configuration on the universal covering space creates integer fluxes. Furthermore, only the fractional part in Eq.~\eqref{wilsonflux} contributes to the phases of the charged fields in terms of the Wilson line values. The integer part characterized by $k_i$ might, however, have interesting consequences in terms of the charged spectrum of particles located at the fixed points, see \cite{Lee:2003mc}. 

These observations fit nicely with the ones made in heterotic string theory compactified on (blowups of) heterotic orbifolds with line bundles. First, in \cite{Nibbelink:2009sp} it was observed that the cycles that resolve the orbifold fixed points to a smooth Calabi-Yau manifold carry flux which is determined by the local orbifold gauge data (i.e.\ the Wilson lines plus a universal, constant flux called gauge shift). Second, in \cite{Blaszczyk:2011ig} a local (non-compact) version of the Hirzebruch-Riemann-Roch (HRR) index theorem was used to show that there can be fractional multiplicities $\pm1/N$ for a $\mathbb{Z}_N$ orbifold such that the contributions sum up to integers for the full (compact) HRR index. They directly correspond to the fractional localized fluxes which also add up to an integer total flux when integrated over the internal space. Matching between the orbifold blowup description in string theory and the geometrical interpretation above is an encouraging consistency check.

Now we generalize our previous results to non-vanishing, constant bulk flux, i.e.\ to a constant field strength $F$ with
\begin{align}
\langle F \rangle = d\langle A \rangle= f v_2 = \text{const} \,.
\end{align}
On the universal covering space this can be described by the gauge field background
\begin{align}
\langle A \rangle = -f y_2 dy_1 \,.
\label{fluxbackground}
\end{align}
The projection to the compact spaces becomes more subtle because the gauge field is a linear function of the coordinates. For a consistent background on $T^2$ we have to impose
\begin{align}
\langle A(y + \lambda) \rangle = \langle A(y) \rangle - d\Lambda(y + \lambda) \quad \text{i.e.,} \quad d \Lambda = (m \lambda_2) f dy_1 \,,
\label{fluxtrafo}
\end{align}
where $\Lambda(y)$ is restricted by Eq.~\eqref{periodgauge}. The gauge field on equivalent points on $\mathbb{R}^2$ is allowed to differ by a gauge transformation that can be projected to $T^2$. On the torus the additional Wilson lines have no direct physical consequence in terms of particle masses. They rather correspond to a certain choice of coordinates \cite{Bachas:1995ik}.\footnote{Furthermore, these configurations do not project to the orbifold.}

Inserting the ansatz in Eq.~\eqref{fluxbackground} and $\lambda = (1,1)$, the gauge parameter has to satisfy
\begin{align}
\Lambda(y + \lambda) - \Lambda(y) = \int_{y}^{y+\lambda} d \Lambda = \int_0^1 f dy_1 = f \in \frac{2\pi}{q} \mathbb{Z} \,.
\end{align}
This yields the appropriate flux quantization on the torus. As required, we find integral periods on closed surfaces for the field strength $\langle F \rangle$,
\begin{align}
\frac{q}{2\pi} \int_{T^2} \langle F \rangle  = \frac{qf}{2\pi} \equiv M \in \mathbb{Z} \,.
\label{sec:wilsonlines:eq:torus_quantization}
\end{align}
The same holds on the orbifold, which due to the reduced area of the fundamental domain leads to\footnote{The minus sign in the definition of $N$ represents left-handed, massless 4d Weyl fermions for $N > 0$, see Sec.~\ref{sec:Anomalies} and \ref{sec:Wavefunctions}.}
\begin{align}
\frac{q}{2\pi} \int_{T^2/\mathbb{Z}_2} \langle F \rangle = \frac{qf}{4\pi} \equiv -N \in \mathbb{Z} \,.
\label{orbquantization}
\end{align}
Hence the flux density $f$ is twice as big compared to $T^2$, see \cite{Braun:2006se}.

For non-vanishing flux a continuous deformation of closed paths in the evaluation of Wilson lines is not possible anymore. However, the discrete Wilson lines on $T^2/\mathbb{Z}_2$ can still be interpreted as singular gauge fluxes on the fixed points and do not lose their relevance in the flux background. In the description of vortices and localized fluxes this can be directly observed, since the field strength $F$ characterizes the different configurations. The corresponding gauge field background on the orbifold is simply the superposition $\langle A \rangle + A^{\text{orb}}$.

\section{6d Supergravity}
\label{sec:6dSugra}

The bosonic part of the six-dimensional supergravity action with a $U(1)$ gauge
field is given by\footnote{We use the differential geometry conventions of \cite{blumenhagen2012basic}.}
\begin{align}
S = \int \left(\frac{1}{2} R - \frac{1}{2}d \phi \wedge \ast d \phi - \frac{1}{2}e^{2\phi} H \wedge \ast H - \frac{1}{2}e^{\phi} F \wedge \ast F \right) \,,
\end{align}
involving the Ricci scalar $R$, the dilaton $\phi$, the field strengths of the gauge field $A = A_M dx^M$ and the antisymmetric tensor field $B = \frac{1}{2} B_{MN} dx^M \wedge dx^N$,
\begin{align}
F = dA \comma H = dB - X^0_3 \,.
\end{align}
Here, we use six-dimensional Planck units $M_{6} = 1$. The 3-form $X^0_3$ is the difference between Chern-Simons forms for the spin connection $\omega$ and the gauge field $A$, 
\begin{align}
X_3^0 &= \omega_{3L} - \omega_{3G} \,,\\ 
\omega_{3L} &= \tr \left( \omega \wedge d \omega + \frac{2}{3} \omega \wedge \omega \wedge \omega \right) \comma \omega_{3G} = A \wedge F \,.
\end{align}
The exterior derivative of the 3-form $X^0_3$ is a gauge invariant 4-form,
\begin{align}
dX_3^0 = X_4 = \tr \left( R \wedge R\right) - F \wedge F \,,
\end{align}
and under a gauge variation
\begin{align}
\delta \omega = d \theta + [\omega, \theta] \comma \delta A = d \Lambda \,,
\end{align}
$X^0_3$ transforms as
\begin{align}
\delta X_3^0 = d \left( \tr (\theta d \omega) - \Lambda F \right) \equiv dX_2^1 \,.
\end{align}
Demanding a gauge invariant field strength $H$, i.e.\ $\delta H = \delta dB - \delta X_3^0 =0 $, one obtains the gauge variation of the antisymmetric tensor field
\begin{align}
\delta B = \tr ( \theta d \omega) - \Lambda F + dC = X^1_2 + dC \,.
\end{align}
Note that the 1-form $C$ parametrizes an additional gauge invariance of the field strength $H$. 

The allowed field content of supergravity in six dimensions is strongly constrained by the requirement that all bulk and fixed point anomalies cancel. The vanishing of the irreducible gravitational anomaly implies the existence of a large number of hypermultiplets. Reducible anomalies can be canceled by the Green-Schwarz mechanism, but the necessary factorization of the anomaly polynomial typically also requires a large number of hypermultiplets charged under the gauge group \cite{Erler:1993zy}. In addition, fixed point anomalies have to be canceled\footnote{We have found some solutions to the anomaly conditions, which will be described in \cite{bdrs1}; a case with several $U(1)$ factors has been analyzed in \cite{Buchmuller:2007qf, Park:2011wv}.}. The focus of this paper is on the cancellation of the four-dimensional chiral anomaly, with contributions both from chiral boundary conditions and gauge flux. For simplicity, in a first step, we therefore ignore
gravitational anomalies. We can then set $\omega_{3L} = 0$ and use $X_3^0 = - A \wedge F$. The complete cancellation of the gravitational anomalies and the influence of the singular curvature located at the fixed points will be discussed in \cite{bdrs1}.

Consider now the orbifold $T^2/\mathbb{Z}_2$ with background flux
\begin{align}
A = \langle A \rangle + \hat{A} \comma F = \langle F \rangle + \hat{F} \comma \text{with} \, \langle F \rangle = d \langle A \rangle \,,
\end{align}
where the hat denotes fluctuations around the background field. In order to preserve four-dimensional Lorentz invariance, the background field can only depend on the coordinates $y$ of the compact space. Neglecting the gravitational backreaction\footnote{This approximation is valid since the contribution to $R_{MN} \propto e^{\phi} f^2/(M_{6}^2 V_2) = e^{\phi} f^2/(M_{\text{P}} \sqrt{V_2})$ is Planck-mass suppressed.}, the equations of motion imply that the flux is constant,
\begin{align}
\langle F \rangle = f v_2 = \tfrac{1}{2} f\epsilon_{mn} dy_m \wedge dy_n \,.
\label{sec:6dSugra:eq:background_flux}
\end{align}
For the field strength $H$ one then obtains \cite{Braun:2006se}
\begin{align}
H = dB - X_0^3 = d \tilde{B} + \hat{A} \wedge \hat{F} + 2 \hat{A} \wedge \langle F \rangle \comma \text{with} \, \tilde{B} = B - \langle A \rangle \wedge \hat{A} \,.
\end{align}
The redefined antisymmetric tensor field can be written as 
\begin{align}\label{dBtilde}
d \tilde{B} = db \wedge v_2 + d \hat{B} \,,
\end{align}
where $b$ is a real scalar field. This yields for the field strength $H$,
\begin{align}
H = \left(db + 2f \hat{A}\right) v_2 + d\hat{B} + \hat{A} \wedge \hat{F} \,.
\end{align}
It is now straightforward to evaluate the gauge part of the action with background flux. The four-dimensional classical effective action of the zero modes is obtained by neglecting the dependence of the fields $\hat{A}, \, b, \, d\hat{B}, \, \phi$, and $r$ on the compact coordinates, and by using $\hat{A}_m = 0$ on the orbifold. The result takes the simple form 
\begin{align}
S_\text{G} = &\int_{M\times X} \Big(-\frac{1}{2} e^{2\phi} H \wedge \ast H
  - \frac{1}{2} e^\phi F
  \wedge \ast F \Big) \nonumber\\
\simeq &\int_M\Big(- \frac{1}{2}s \hat{F}\wedge\ast\hat{F} 
- \frac{f^2}{2t^2 s} \nonumber\\
&\hspace*{1cm}-\frac{1}{2t^2} (db + 2f \hat{A})\wedge\ast
(db + 2f \hat{A})\nonumber\\
&\hspace*{1cm}-\frac{s^2}{2} (d\hat{B} +
  \hat{A}\wedge\hat{F})\wedge\ast (d\hat{B} + \hat{A}\wedge \hat{F})\Big)\,, \label{SG1}
\end{align}
where we have replaced radion and dilaton by the real scalar fields
\begin{align}
t = r^2 e^{-\phi}\,, \quad s = r^2 e^\phi\,.
\end{align}
Further, we switch to four-dimensional Planck units $M_{\text{P}}^2 = (M_{6})^4 V_{2} = 1$, with $V_{2}$ the physical size of the internal space\footnote{In this way the factor $1/2$ of the orbifold volume is absorbed in the effective Planck mass.}. Note that $S_G$ is invariant under the gauge transformation
\begin{align}
\delta \hat{A} = d\Lambda\,, \quad \delta b = - 2f\Lambda\,,
\quad \delta \hat{B} = - \Lambda F \,.
\label{axiontransform}
\end{align}

The action \eqref{SG1}  describes an interacting vector boson, which becomes massive by absorbing the axion $b$, corresponding to the field redefinition $\hat{A} \rightarrow \hat{A} -\tfrac{1}{2f}db$. The mass of the vector boson depends on the vacuum expectation value of the modulus fields $s$ and $t$ and can be read off from Eq.~\eqref{SG1},
\begin{align}
m_{\hat{A}}^2 = \frac{4f^2}{s_0 t_0^2}\,.
\end{align}
Contrary to the standard form of the Stueckelberg mechanism the mixing between axion and vector field is not generated by the Green-Schwarz term but by the classical flux $\langle F \rangle$. Hence the origin of the mass is the classical self-interaction of the gauge field, encoded in the Chern-Simons term, and the background flux.

\section{Anomalies and Green-Schwarz mechanism}
\label{sec:Anomalies}

We now turn to the second source of the vector boson mass, the axion-vector boson mixing due to the anomaly. Let us first recall the anomalies associated to a 6d Weyl fermion $\psi$ with $U(1)$ charge $q$,
\begin{align}\label{fermionR}
\mathcal{L}_f = \bar{\psi}(x) i\Gamma^a e_a^M D_M \psi(x) \,, \quad \Gamma^7
\psi = -\psi\,.
\end{align}
Here, $\Gamma^0, \dots, \Gamma^6$ are the $\Gamma$ matrices in six dimensions, $e_a^M$ the inverse vielbein, and $\Gamma^7 = \Gamma^0\cdot \ldots \cdot\Gamma^6$. The gauge covariant derivative is $D_M = \partial_M + i q A_M$. The 6d Weyl fermion $\psi$ contains two 4d Weyl fermions of opposite chirality, $\psi = (\psi_L,\psi_R)$, with $\gamma_5 \psi_L = -\psi_L$ and $\gamma_5 \psi_R = \psi_R$. We impose chiral boundary conditions,
\begin{align}
\psi_L(x^\mu,y_m) = \psi_L(x^\mu,-y_m) \,, \quad \psi_R(x^\mu,y_m) =
-\psi_R(x^\mu,-y_m) \,,
\label{sec:Anomalies:eq:chiral_bc}
\end{align}
which correspond to one possible embedding of the orbifold twist into the $SU(2)_R$ symmetry of the Lagrangian~\eqref{fermionR}.

\subsection{Bulk and fixed point anomalies}

The chirality of the Weyl fermion leads to an anomaly, i.e.~the effective action derived from Eq.~\eqref{fermionR}
\begin{align}
\Gamma_f[A] = -i \ln{\int D\psi D\bar{\psi} \exp{\left(i\int \mathcal{L}_f\right)}}
\end{align}
is not invariant under gauge transformations $\delta A = d\Lambda$,
\begin{align}
\delta\Gamma_f[A] = -\int \mathcal{A} \,.
\end{align}
The anomaly $\mathcal{A}$ contains a familiar bulk term \cite{Erler:1993zy} and, due to the boundary conditions, additional contributions which are localized at the fixed points \cite{Scrucca:2004jn,vonGersdorff:2006nt}. An explicit calculation yields \cite{Asaka:2002my}
\begin{align}\label{6Danomaly}
\mathcal{A} = \Lambda F \wedge \left(\frac{\beta}{2} F\wedge F +
  \alpha \delta_O F\wedge v_2 \right)\,.
\end{align}
Here $\beta = -q^4/(2\pi)^3$, $\alpha = q^3/(2\pi)^2$, and
$\delta_O(y)$, accounts for the localized contributions to the anomaly, which are equally distributed over the four orbifold fixed points $\zeta_1,\dots, \zeta_4$,
\begin{align}
\delta_O(y) = \frac{1}{4}\sum_{i=1}^4 \delta (y-\zeta_i) \,.
\end{align}
This expression is valid for pure chiral boundary conditions without further Wilson lines. It can be explicitly derived from the fermionic wave functions in the case without flux, see \cite{Asaka:2002my}. Nevertheless, a more general form of the anomaly similar to Eq.~\eqref{6Danomaly} can be found for arbitrary Wilson line backgrounds by inserting the singular flux configurations discussed in Eq.~\eqref{wilsonflux}. The Wilson lines, however, do not alter the number of fermionic zero modes generated by the flux. They only affect the zero mode arising due to the chiral boundary conditions. In the following we restrict the discussion of the anomaly to the case without Wilson lines. The completely general treatment will be subject of further investigations \cite{bdrs1}. The six-dimensional anomaly contains an anomaly for the effective theory in four dimensions which is obtained by integrating over the internal space. Without flux, i.e.\ $F = \hat{F}$, the bulk term vanishes, and the anomaly exclusively arises due to the fixed point contributions. In our case the four-dimensional anomaly reads
\begin{align}\label{4Danomaly}
\mathcal{A}_4 = \alpha \int_X \delta_O \Lambda \hat{F}\wedge \hat{F}\wedge v_2
= \alpha \Lambda \hat{F} \wedge \hat{F} \,.
\end{align}
This is precisely the chiral anomaly of a single four-dimensional Weyl fermion, which corresponds to the zero mode of the six-dimensional Weyl fermion obeying chiral boundary conditions.

Given the form of the anomaly \eqref{6Danomaly}, one can immediately write down the Green-Schwarz counter term which cancels this anomaly,
\begin{align}\label{GS1}
S_\text{GS}[A,B] = -\int \left(\frac{\beta}{2} A\wedge F + \alpha
  \delta_O A\wedge v_2\right) \wedge dB\,.
\end{align}
Using the transformation property $\delta dB = -d\Lambda\wedge F$ and performing an integration by parts, one obtains 
\begin{align}
\delta S_\text{GS} =\int \left( \frac{\beta}{2} F\wedge F +
  \alpha \delta_O F\wedge v_2\right) \wedge \Lambda F
= -\delta\Gamma_f \,,
\label{anomaly}
\end{align}
i.e., the sum $\Gamma_f[A] + S_\text{GS}[A,B]$ is indeed gauge invariant.

\subsection{Effective action for gauge field and axions}
We can now easily obtain the complete four-dimensional action for the zero modes in the presence of the background flux. Using Eqs.~\eqref{dBtilde}, \eqref{SG1} and \eqref{GS1} one finds \newpage
\begin{align}
S_\text{G} + S_\text{GS} = \int_M\Big(&-\frac{1}{2}s \hat{F}\wedge\ast\hat{F} 
- \frac{f^2}{2t^2 s}  \nonumber\\
&-\frac{1}{2t^2} (db + 2f \hat{A})\wedge\ast
(db + 2f \hat{A})\nonumber\\ 
&-\frac{s^2}{2} (d\hat{B} +
  \hat{A}\wedge\hat{F})\wedge\ast (d\hat{B} + \hat{A}\wedge \hat{F}) \nonumber\\ \displaybreak[1]
&-\frac{\beta}{2}\hat{A}\wedge \hat{F} \wedge db - \Big(\alpha +
  \frac{\beta}{2}f\Big)\hat{A}\wedge d\hat{B} \Big) \,.
\end{align}
Introducing $\hat{H} = d\hat{B} + \hat{A}\wedge \hat{F}$ and a Lagrange multiplier field $c$ enforcing that $\hat{H} - \hat{A}\wedge \hat{F}$ is closed,
\begin{align}
\Delta S_\text{G} = \int_M c\ d(\hat{H} - \hat{A}\wedge \hat{F})\,, 
\end{align}
one obtains after integration by parts
\begin{align}
S_\text{G} + S_\text{GS} = \int_M\Big(&-\frac{1}{2} s\hat{F}\wedge\ast\hat{F} 
- \frac{f^2}{2t^2 s}  \nonumber\\
&-\frac{1}{2t^2} (db + 2f \hat{A})\wedge\ast
(db + 2f \hat{A})\nonumber\\
&-\frac{s^2}{2} \hat{H}\wedge\ast \hat{H} -
\Big(\Big(\alpha +\frac{\beta}{2}f\Big)\hat{A} + dc\Big)\wedge \hat{H} \nonumber\\
&-\hat{A}\wedge\hat{F}\wedge\Big(\frac{\beta}{2} db + dc\Big) \Big)\,.
\end{align}
The field $\hat{H}$ can now be eliminated by its equation of motion,
\begin{align}\label{Heom}
\ast\hat{H} = \frac{1}{s^2} \Big(dc + \Big(\alpha + \frac{\beta}{2}f\Big)
\hat{A} \Big) \,,
\end{align}
which yields the final result
\begin{align}\label{SAbc}
S_\text{G} + S_\text{GS} = \int_M\Big(&-\frac{1}{2} s\hat{F}\wedge\ast\hat{F} 
- \frac{f^2}{2t^2 s}  \nonumber\\
&-\frac{1}{2t^2} (db + 2f \hat{A})\wedge\ast
(db + 2f \hat{A})\nonumber\\
&-\frac{1}{2s^2} \Big(dc +\Big(\alpha +\frac{\beta}{2}f\Big)\hat{A} \Big)\wedge \ast
\Big(dc +\Big(\alpha +\frac{\beta}{2}f\Big)\hat{A} \Big) \nonumber\\
&-\hat{A}\wedge\hat{F}\wedge\Big(\frac{\beta}{2} db + dc\Big) \Big) \,.
\end{align}
Note that the total action is invariant under gauge transformations. Eq.~\eqref{Heom} implies 
\begin{align}
\delta c = - \Big(\alpha + \frac{\beta}{2}f\Big)\Lambda\,.
\label{ctransform}
\end{align}

We are now ready to discuss the cancellation of the 4d anomaly. The contributions from both, flux and orbifold projection, are obtained by inserting $F = \langle F \rangle + \hat{F}$ into Eq.~\eqref{anomaly}, which yields
\begin{align}
\delta \Gamma_f = - \Big(\alpha + \frac{3}{2}\beta f\Big)\int_M
\Lambda \hat{F}\wedge \hat{F} \,.
\end{align}
Note that in addition to the fixed point contribution \eqref{4Danomaly} there is now a second term proportional to $\beta f$, reflecting the additional contribution form zero modes due to the index theorem. Using Eq.~\eqref{ctransform} and $\delta b= -2f\Lambda$ one easily verifies that
\begin{align}
\delta \Gamma_f + \delta(S_\text{G} + S_\text{GS}) = 0 \,.
\end{align}   
It is very satisfactory that the Green-Schwarz term designed to cancel bulk and fixed point anomalies automatically cancels the additional anomaly resulting from the flux by modifying the transformation properties of both axions (cf.\ Eqs.~\eqref{axiontransform}, \eqref{ctransform}).

The action~\eqref{SAbc} contains two ``axions'', $b$ and $c$. Consider first the case without flux, i.e.\ $f=0$. The field redefinition $\hat{A} \rightarrow \hat{A} - \alpha^{-1}dc$ then turns the kinetic term of $c$ into a mass term for $\hat{A}$ and, using
\begin{align}
\Gamma_f[\hat{A} - \alpha^{-1}dc] - \Gamma_f[\hat{A}] =
\int_M c\ \hat{F}\wedge \hat{F} = - \int_M dc\wedge \hat{A}\wedge \hat{F} \,,  
\end{align}
the coupling of $c$ to the Chern-Simons term in Eq.~\eqref{SAbc} is removed. From Eq.~\eqref{SAbc} we then obtain an action for a massive vector field and an axion $b$,
\begin{align}
S_b = 
\int_M\Big(&-\frac{1}{2} s\hat{F}\wedge\ast\hat{F} -\frac{1}{2s^2}
\alpha^2 \hat{A}\wedge \ast\hat{A}
  \nonumber\\
&-\frac{1}{2t^2} db\wedge\ast db + \frac{\beta}{2} b\ \hat{F}\wedge \hat{F} \Big)\,,
\end{align}
which yields for the mass of the vector boson 
\begin{align}
m_{\hat{A}}^2 = \frac{1}{s_0^3}\alpha^2\,.
\label{m2_annom}
\end{align}
This is the standard Stueckelberg mechanism for generating a vector boson mass. The axion $b$ is massless and couples to $\hat{F}\wedge \hat{F}$ in the familiar way.

The general case with non-zero flux is more complicated. One linear combination, $\chi$, of $b$ and $c$ gives mass to the vector boson, and a second linear combination, $a$, of $b$ and $c$ plays the role of a massless axion. Since the flux is quantized, $f = -4\pi N/q$ with $N$ integer\footnote{The minus sign corresponds to left-handed zero modes in 4d for $N>0$, see Sec.~\ref{sec:Wavefunctions}.}, and $\beta = -q\alpha/(2\pi)$, the coefficient $\alpha + \beta f/2 = \alpha (N+1)$ counts the total number of chiral fermions. 

The linear combination $\chi$  is determined by performing a field redefinition, $\hat{A} \rightarrow \hat{A} + d\chi$, and demanding that all mixing terms between $db$, $dc$, $d\chi$ and $\hat{A}$ vanish. This gives 
\begin{align}\label{chi}
\chi = - \frac{2 f s_0^2 b + \alpha (1+N) t_0^2 c}{4 f^2 s_0^2 +\alpha^2 (N+1)^2 t_0^2} \,.
\end{align} 
After this field redefinition, the kinetic terms for $b$ and $c$ yield a mass term for the vector field and a kinetic term for the linear combination 
\begin{align}\label{a}
a = 2f c - \alpha (N+1) b \,.
\end{align}
From Eqs.~\eqref{SAbc}, \eqref{chi} and \eqref{a} one obtains
\begin{align}\label{SAa}
S_a =
\int_M\Big(&-\frac{1}{2}s_0 \hat{F}\wedge\ast\hat{F} 
- \frac{f^2}{2t_0^2 s_0}  - \frac{s_0}{2} m_{\hat{A}}^2 \hat{A}\wedge \ast \hat{A}   \nonumber\\
&-\frac{\kappa}{2} da\wedge\ast da + \lambda a \hat{F}\wedge \hat{F}\Big)\,,
\end{align}
where
\begin{align}\label{vectormass}
m_{\hat{A}}^2 &= \frac{4f^2}{s_0 t_0^2} + \frac{1}{s_0^3} \alpha^2(N+1)^2 \,, \\
\kappa &= \frac{1}{4 f^2 s_0^2 + \alpha^2 (N+1)^2 t_0^2}\,, \\
\lambda & = \frac{2 f s_0^2 - \alpha \beta (N+1) t_0^2}{4 f^2 s_0^2 + \alpha^2 (N+1)^2 t_0^2} \,.
\end{align}
The mass formula \eqref{vectormass} clearly shows the two contributions to the vector boson mass from the classical flux and the anomaly due to the total number of $(N+1)$ chiral zero modes. Restoring dimensionful parameters, the vector boson mass is
\begin{equation}
m_{\hat A}^2 = m^2_\mathrm{class} + m^2_\mathrm{anom} = \frac{1}{M_P V_2^{3/2}} \left( e^{\phi_0} N^2 \frac{64 \pi^2 }{q^2} + e^{-3\phi_0} (N+1)^2 \frac{ q^6}{4 \pi^2}\right).
\end{equation}
Both contributions scale in the same way with the 4d Planck mass and the volume of the internal space whereas the dependence on the dilaton field is different.
In the case without flux, \(N = 0\), we recover Eq.~\eqref{m2_annom} with appropriate dimensionful parameters.
The axion coupling strength and the normalization of its kinetic term are determined by the vacuum expectation values of the moduli.

The mechanism to generate a vector boson mass discussed in this section is similar to observations made in heterotic string theory compactified on 6d orbifolds. In \cite{Ludeling:2012cu} it was discussed that a $U(1)$ vector boson can be massive even if it is non-anomalous: the mass simply arises from its coupling to the axion. Furthermore, \cite{GrootNibbelink:2007ew,Blaszczyk:2011ig} studied anomaly cancellation on orbifolds and their smooth Calabi-Yau resolutions with flux. On the orbifold without flux all anomalies are canceled via the universal axion $c$ which is the partner of the dilaton. Going to the smooth blowup with flux, it was found that the anomalies are canceled by a linear combination of the universal axion and other axions which are the partners of the K\"ahler moduli of the cycles that carry the flux. In our setup they correspond to the field $b$.

\section{Wave functions}
\label{sec:Wavefunctions}

In this section we first briefly recall the bosonic and fermionic mass spectra, both with and without bulk flux. In the case with flux fermionic chiral zero modes arise and the mass spectrum is independent of shape moduli and Wilson lines. Nevertheless, the orbifold Wilson lines influence the shape of the wave functions in a specific way. We construct explicit expressions for the wave functions in a flux background on \(T^2\) and their projection to \(T^2/\mathbb Z_2\). Moreover, we explicitly demonstrate the matching of the fermionic zero modes predicted by the index theorem and the number of linear independent wave functions.

\subsection{Mass spectra of the Dirac and Laplace operators}
\label{sec:wavefunctions:spectra}
Without bulk flux the scalar mass spectrum on $T^2$ including Wilson lines has been studied before in \cite{Ghilencea:2005vm}. Implementing the Wilson lines as modified periodicity conditions for charged fields, see Eq.~\eqref{periodicitymod}, the wave functions in the compact dimensions can be decomposed as
\begin{align}
\phi(y) = e^{i q (\alpha_1 y_1 + \alpha_2 y_2)} \sum_{r,s \in \mathbb Z} a_{rs} e^{2 \pi i (r y_1 + s y_2)} \equiv e^{i q (\alpha_1 y_1 + \alpha_2 y_2)} \sum_{r,s \in \mathbb Z} \phi_{rs}(y) \,.
\end{align}
Plugging a mode $\phi_{rs}$ into the Laplace equation of the internal space we obtain the Kaluza-Klein masses
\begin{equation}
(g_2^{mn} \partial_m \partial_n) \phi_{rs} \equiv	\Delta_2 \phi_{rs} = - m_{rs}^2 \phi_{rs} \,.
	\label{sec:wavefunctions:eq:spectrum_no_flux}
\end{equation}
For vanishing bulk flux they depend on the Wilson lines and shape moduli $\tau$,
\begin{align}
m_{rs}^2 = \frac{1}{\tau_2} \vert (2 \pi s + q \alpha_2) - \tau (2 \pi r + q \alpha_1) \vert^2\,.
\end{align}
Restoring the physical dimension, the size of the internal space enters the Kaluza-Klein mass formula,
\begin{align}
m_{\text{phys}}^2 = \frac{m^2}{V_2}  \,.
\label{physicalmass}
\end{align}
The fermionic mass spectrum is identical and supersymmetry is unbroken.

For non-vanishing bulk flux the mass spectrum changes dramatically \cite{Bachas:1995ik}. We choose the gauge field background of Sec.~\ref{subsec:WilsonWithout},
\begin{align}
	\langle A \rangle = -f y_2 \, dy_1\,, \quad \langle F \rangle = f v_2 \,,
	\label{sec:wavefunctions:eq:gauge_with_flux}
\end{align}
where $\langle F \rangle$ is subject to the quantization condition \eqref{sec:wilsonlines:eq:torus_quantization} on $T^2$, which requires $fq/ 2 \pi \equiv M \in \mathbb{Z}$. Due to the non-trivial behavior of $A$ under translations it is more subtle to find a solution for the equations of motion. The explicit expression for the fermionic zero modes is presented in the following section.
 
There is, however, a shortcut to obtain the bosonic and fermionic mass spectrum. The commutator of covariant derivatives does not vanish in the gauge background~\eqref{sec:wavefunctions:eq:gauge_with_flux}. As a result, a treatment analogous to the harmonic oscillator is possible \cite{Bachas:1995ik} and yields for charged, bosonic fields
\begin{align}
	m^2_n = 4 \pi |M| \left( n + \frac{1}{2} \right)\,.
\end{align}
The relation to dimensionful quantities is again given by Eq.~\eqref{physicalmass}. Hence, the structure of the mass spectrum in the flux background fundamentally differs from Eq.~\eqref{sec:wavefunctions:eq:spectrum_no_flux}. Most importantly, it is independent of the shape moduli and Wilson lines. Hence, a stabilization of shape moduli in the flux background via the Casimir energy, as in \cite{Ghilencea:2005vm, Buchmuller:2009er}, seems to be problematic.

In a similar vein, we follow \cite{Braun:2006se, Bachas:1995ik} to find the fermionic Kaluza-Klein masses. We decompose the charged 6d Weyl fermion into a tensor product of a 4d and a 2d Weyl fermion of opposite chiralities \cite{Abe:2013bca},
\begin{align}
\psi(x^{\mu},y_m) = \psi_{4L}(x^{\mu}) \otimes \psi_{2R}(y_m) + \psi_{4R}(x^{\mu}) \otimes \psi_{2L}(y_m) \,.
\end{align}
The fermionic masses are determined by the eigenvalues of the squared Dirac operator in the internal space. It acts on charged 6d Weyl fermions as
\begin{align}
(\Gamma^m D_m)^2 = (g_2)^{mn} D_m D_n - \frac{1}{2} \Gamma^m \Gamma^n [D_n, D_m] = \Delta_2 - 2 \pi M \gamma_5 \Gamma^7 \,.
\end{align}
Compared to the bosonic spectrum the fermion masses get shifted depending on their 4d chirality
\begin{equation}
	m_n^2 = 4 \pi |M| \left( n + \frac{1}{2} \mp \frac{1}{2} \right)\,.
\end{equation}
The upper sign holds for the left-handed 4d components $\psi_{4L}$.

Therefore, the flux gives rise to left-handed zero modes in four dimensions. As shown in Sec.~\ref{sec:Anomalies}, these contribute to the 4d anomaly. Finally, we want to point out that charged scalar fields always have a mass $m^2 \geq 2 \pi |M|$. Therefore, supersymmetry is spontaneously broken at the scale of compactification.

The mass spectra remain valid on $T^2/\mathbb{Z}_2$ up to proper flux quantization, cf.~Eqs.~\eqref{sec:wilsonlines:eq:torus_quantization} and~\eqref{orbquantization}.

\subsection{Zero mode wave functions with Wilson lines}
\label{sec:wavefunction:expression}

To find explicit expressions for the wave functions of fermionic zero modes on $T^2$ for non-vanishing bulk flux, we follow \cite{Cremades:2004wa}. However, we use the coordinates introduced in Sec.~\ref{sec:InternalSpace}. They allow us to pick the particularly simple gauge given in Eq.~\eqref{sec:wavefunctions:eq:gauge_with_flux}.

For a charged field this background implies a gauge transformation associated with torus translations, see Eq.~\eqref{fluxtrafo},
\begin{align}
\Lambda = f y_1 = \frac{2\pi}{q} M y_1 \,.
\label{eq:orbgauge}
\end{align}
Moreover, we absorb the phases associated to Wilson lines into modified boundary conditions. For a field of charge $q$ these are
\begin{equation}
\begin{aligned}
	\phi(y + \lambda_1) &= e^{iq\alpha_1} \phi(y),\\
	\phi(y + \lambda_2) &= e^{iq(\alpha_2+f y_1)} \phi(y)\,.
\end{aligned}
\label{sec:BulkField:eq:Periodicities}
\end{equation}
While $\alpha_m$ correspond to continuous Wilson lines on $T^2$, their values are restricted to be $k_m \pi/q$ with $k_m \in \{0,1\}$ on $T^2/\mathbb{Z}_2$, see Eq.~\eqref{translationphase}. These conditions are satisfied by the ansatz
\begin{equation}
\phi(y) = e^{i q (\alpha_1 y_1 + \alpha_2 y_2)} \sum_n f_n(y_2) e^{2 \pi i n y_1}.
	\label{sec:BulkField:eq:AnsatzPsi}
\end{equation}
The functions \(f_n(y_2)\) only depend on the summation index \(n\) and \(y_2\), but not on \(y_1\).
They fulfill the recurrence relation
\begin{equation}
	f_n(y_2 + 1) = f_{n - M}(y_2)\,,
	\label{sec:BulkField:eq:RecurrenceRelation}
\end{equation}
due to the associated gauge transformation, Eq.~\eqref{eq:orbgauge}.

The Dirac equation in the internal space reads
\begin{align}
i \Gamma^m D_m \begin{pmatrix} \psi_{2L} \\ \psi_{2R} \end{pmatrix} = \begin{pmatrix} 0 & \tau D_1 - D_2\\ -\bar{\tau} D_1 + D_2 & 0 \end{pmatrix} \begin{pmatrix} \psi_{2L} \\ \psi_{2R} \end{pmatrix} = 0\,.
\end{align}
It acts on the 2d component of the Weyl fermion.
In order to obtain left-handed zero modes in 4d we need normalizable zero modes for $\psi_{2R}$, i.e.~$M < 0$. There is a set of linear independent wave functions labeled by $j \in \{ 0, \dots, |M|-1\}$,
\begin{align}
	\psi^j &= \mathcal N e^{ i q \alpha_1(y_1 + \tau y_2) - i \pi M \tau y_2^2} \, \theta\!\begin{bmatrix} j/M \\ -\tfrac{q}{2\pi} (\alpha_1 \tau - \alpha_2) \end{bmatrix}\!\big( M ( y_1 + \tau y_2 ), - M \tau\big)\,,
	\label{sec:wavefunctions:eq:psi+final}
\end{align}
where \(\mathcal N\) is a normalization factor. Further, we use the Jacobi theta function \cite{polchinski1998string}
\begin{equation}
	\theta\!\begin{bmatrix} \alpha \\ \beta \end{bmatrix}\! (x, y) = \sum_n e^{\pi i y (n + \alpha)^2} e^{2 \pi i  (n + \alpha) (x + \beta) }\,.
\end{equation}
Our expressions match those of the torus wave functions in \cite{Cremades:2004wa,Bachas:1995ik}. 

To project the wave functions to $T^2/\mathbb Z_2$, we construct linear combinations with well-defined transformation properties under the action of $P$, see Sec.~\ref{subsec:WilsonWithout} and \cite{Braun:2006se}.
Moreover, the flux quantization condition changes.
Therefore, we introduce $-N = M/2$, the number of orbifold flux quanta\footnote{Note that our flux quantization conditions on $T^2/\mathbb{Z}_2$ differ from those in \cite{Abe:2013bca}, which leads to a different number of fermionic zero modes.} defined in Eq.~\eqref{orbquantization}.

The linear combinations $\psi^j_+ = (1/\sqrt{2}) (\psi^j_{2R}(y) + \psi^j_{2R}(-y))$ transforming even under $P$ can be expressed as
\begin{multline}
	\psi^j_+(y; k_m) = \mathcal N' \, e^{2\pi i N \tau  y_2^2 } \sum_{n \in \mathbb Z} e^{2 \pi i \tau N \left( n - \frac{j}{2N} \right)^2 - i \pi \left( n - \frac{j}{2N}\right) (k_1 \tau - k_2)} \\
	\times 	 \cos\left[ 2 \pi \left( -2nN + j + \frac{k_1}{2}\right) (y_1 + \tau y_2) \right]\,;
	\label{sec:wavefunctions:eq:orbifold_wave}
\end{multline}
here \(\mathcal N'\) is an adjusted normalization factor. For the $P$-odd combinations, one has to replace the cosine with a sine.

In the case without bulk flux and Wilson lines we expect a single fermionic zero mode arising due to the chiral boundary condition on the orbifold. Indeed, the constant wave function, see Fig.~\ref{sec:wavefunctions:fig:wavefunctions_comparison_m0}, solves the equation of motion with vanishing mass. In the presence of Wilson lines around the orbifold fixed points this zero mode disappears, since the constant wave function does not satisfy the modified boundary conditions. Even though the shape of this wave function changes for non-vanishing bulk flux, the considerations above remain valid. Moreover, the index theorem predicts $N$ additional zero modes corresponding to the number of flux quanta. These are present independently of the Wilson line background. Thus, in the absence or presence of localized fluxes we expect to find $(N+1)$ or $N$ independent zero mode wave functions, respectively. Note that the localized fluxes influence the shape of the zero modes. In particular, the wave functions vanish at fixed points with localized flux. It is apparent that the description in terms of orbifold 1-cycles significantly simplifies the treatment of the different wave function profiles. The gauge field background modifies the regions in the bulk where the wave functions are concentrated. This has important consequences for phenomenology affecting e.g.\ interaction terms, see \cite{Cremades:2004wa, bdrs2}.

\begin{figure}
	\begin{center}
	\subfloat[No flux \label{sec:wavefunctions:fig:wavefunctions_comparison_m0}]{
		\includegraphics[height = .25 \textheight]{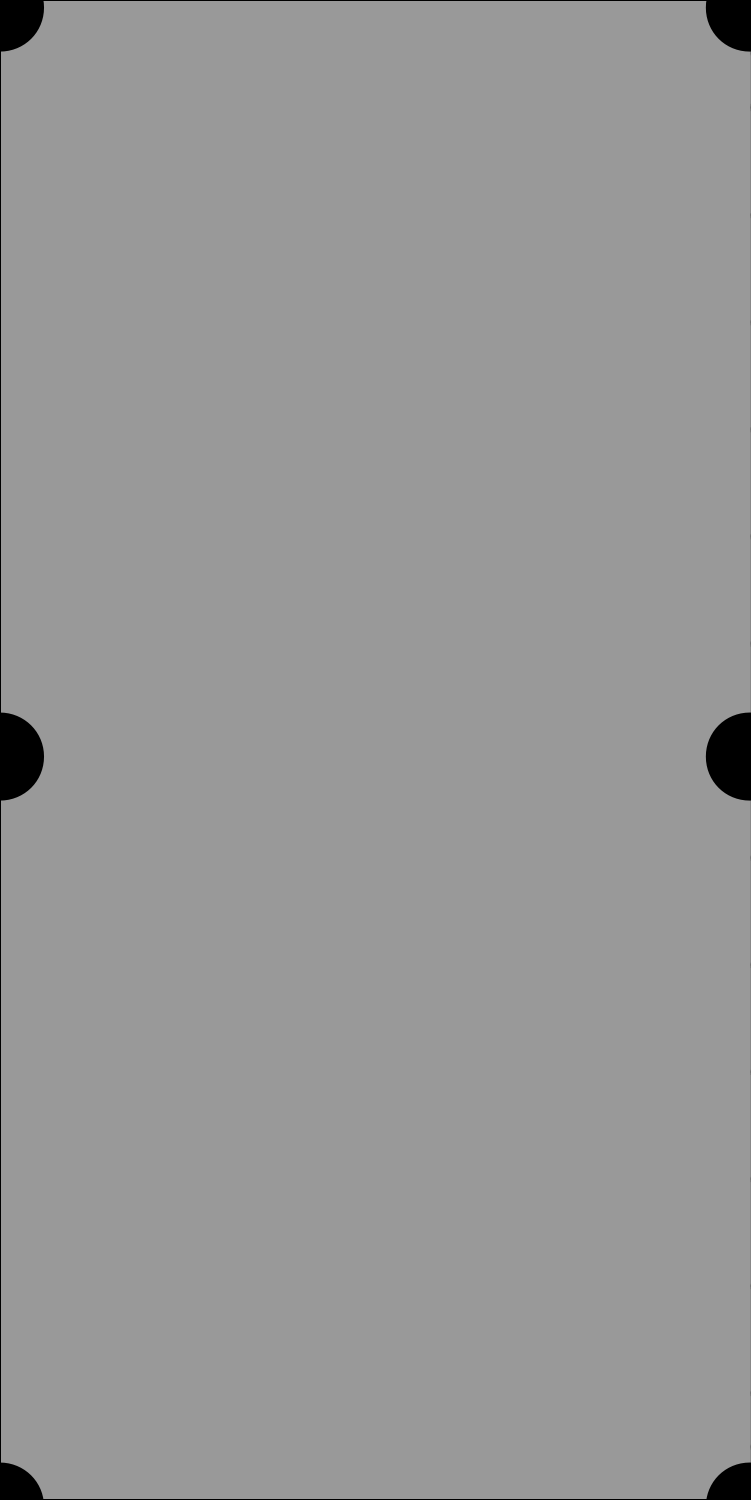}}\hspace*{1.0cm}
	\subfloat[\(N = 2, j = 0\) \label{sec:wavefunctions:fig:wavefunctions_comparison_m2_0}]{
		\includegraphics[height = .25 \textheight]{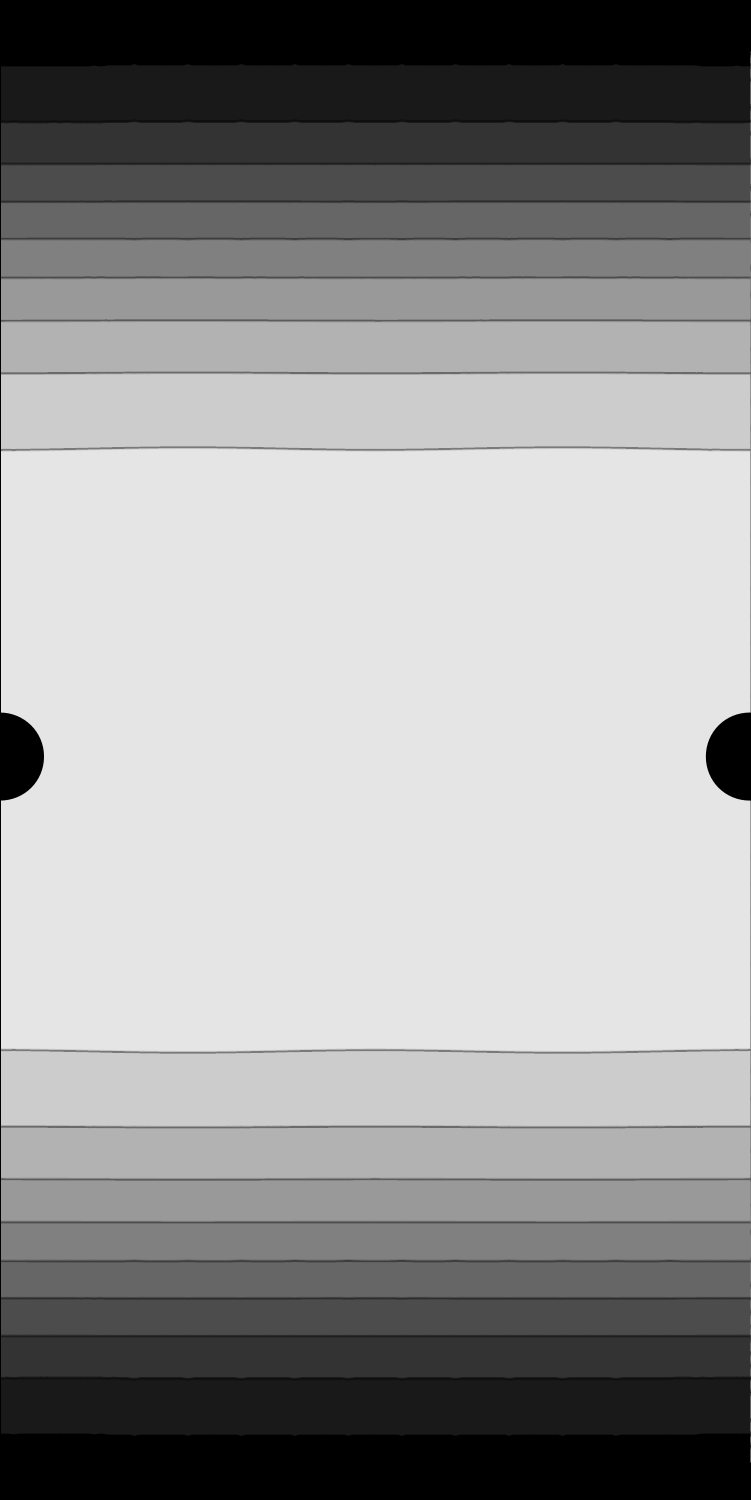}}\hspace*{0.1cm}
	\subfloat[\(N = 2, j = 1\) \label{sec:wavefunctions:fig:wavefunctions_comparison_m2_1}]{
		\includegraphics[height = .25 \textheight]{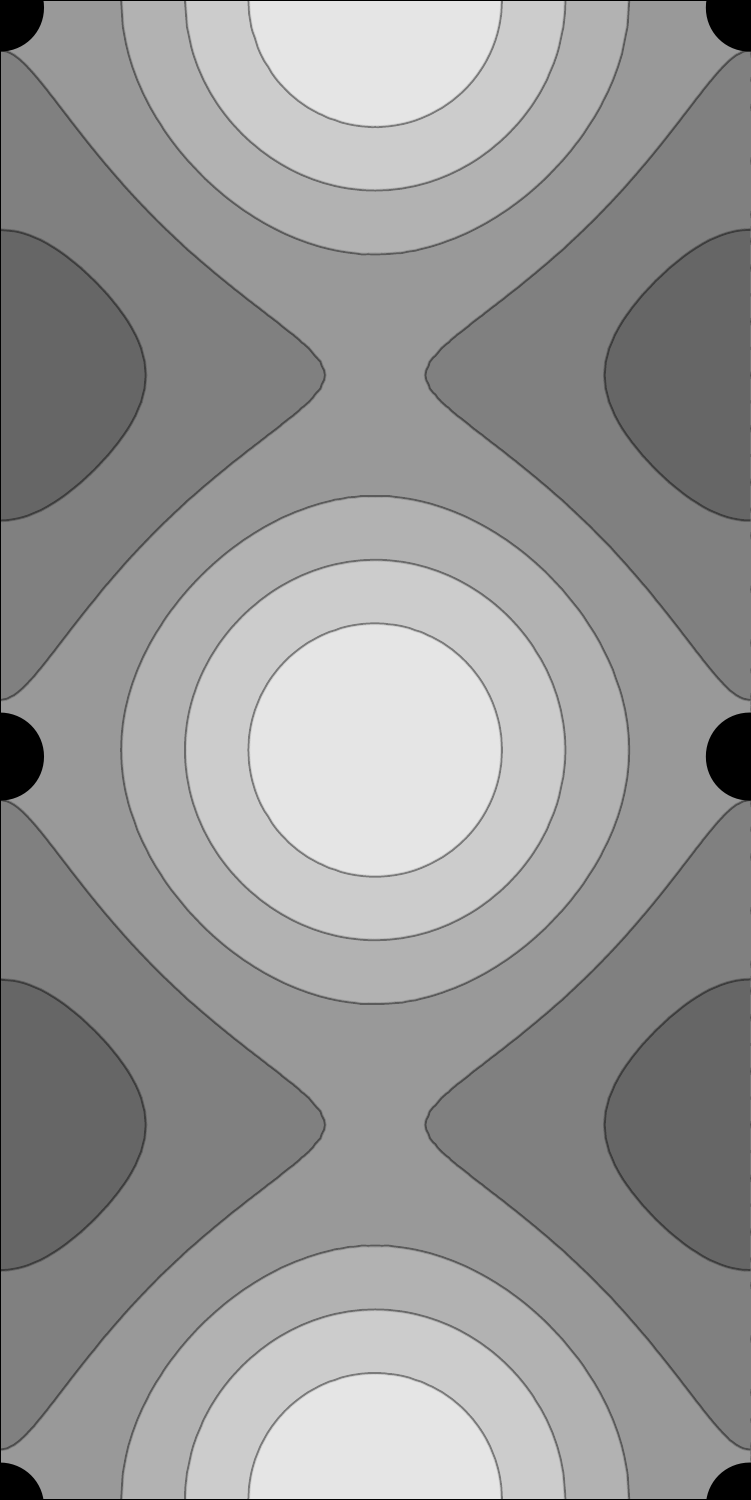}}\hspace*{0.1cm}
	\subfloat[\(N = 2, j = 2\) \label{sec:wavefunctions:fig:wavefunctions_comparison_m2_2}]{
		\includegraphics[height = .25 \textheight]{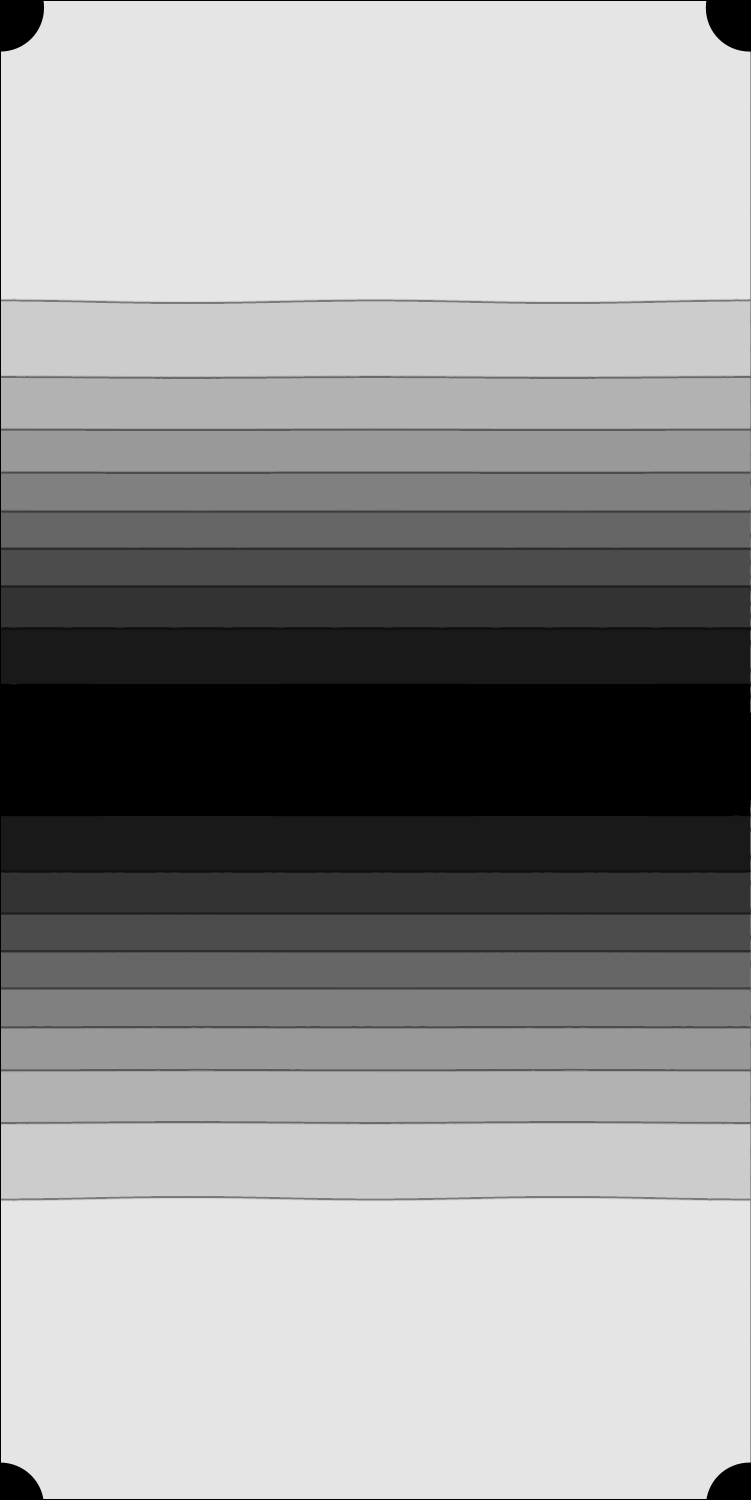}}\hspace*{0.1cm}
	\subfloat{\includegraphics[height = .25 \textheight]{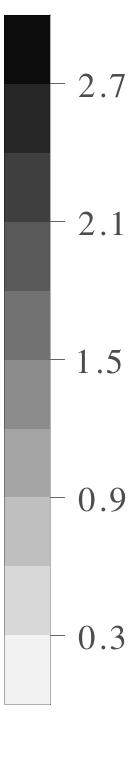}}
	\caption{The (absolute square of the) $P$-even, constant wave function for $N = 0$ (a), and the linear independent wave functions for \(N = 2\) and \(k_m = 0\), (b) to (d). The scale is chosen such that the wave functions integrate to $1/2$ on the orbifold.}
	\label{sec:wavefunctions:fig:wavefunctions_comparison}
	\end{center}
\end{figure}
Consider now the configuration without Wilson lines, $k_1 = k_2 = 0$. Manipulation of the sum in Eq.~\eqref{sec:wavefunctions:eq:orbifold_wave} reveals that $\psi^j_+$ and $\psi^{2N - j}_+$ are identified on the orbifold, i.e., linear independent wave functions are labeled by $j \in \{0, \dots ,N\}$. Hence, there are \((N+1)\) distinct, $P$-even zero mode solutions of the equations of motion. These are accompanied by \((N-1)\) $P$-odd zero modes that are projected out by the chiral boundary conditions~\eqref{sec:Anomalies:eq:chiral_bc}. This is in agreement with the number of zero modes found in the calculation of the anomaly in Sec.~\ref{sec:Anomalies}. As an example, Fig.~\ref{sec:wavefunctions:fig:wavefunctions_comparison} shows the three independent, even wave functions on $T^2/\mathbb{Z}_2$ for $N=2$ without Wilson lines.

In the case of non-zero Wilson lines there are eight possible configurations (cf. Tab.~\ref{Tab:Phases}). Since we restrict our considerations to wave functions satisfying chiral boundary conditions, these are reduced to four configurations. For $N=2$ the three cases with non-trivial Wilson lines are depicted in Fig.~\ref{sec:wavefunctions:fig:wilson_lines}. There, empty circles mark fixed points at which the wave function vanishes. Their patterns agree with our expectation from the decomposition of torus Wilson lines into canonical Wilson lines on $T^2/\mathbb{Z}_2$ according to Tab.~\ref{Tab:Phases}. For example, in the case $k_1 = 1, k_2 = 0$, corresponding to $(P, T_1, T_2) = (+,-,+)$ in the table, the wave function vanishes at $\zeta_2$ and $\zeta_4$ where fluxes are localized, i.e.~$(W_1, W_2, W_3) = (+,-,+)$.  The same holds true for the other configurations of localized fluxes in Tab.~\ref{Tab:Phases}. This correspondence nicely illustrates the benefit of the canonical 1-cycle basis.
\begin{figure}[t]
	\begin{center}
	\subfloat[\(k_1 = 1, k_2 = 0\)]{\includegraphics[height = .25 \textheight]{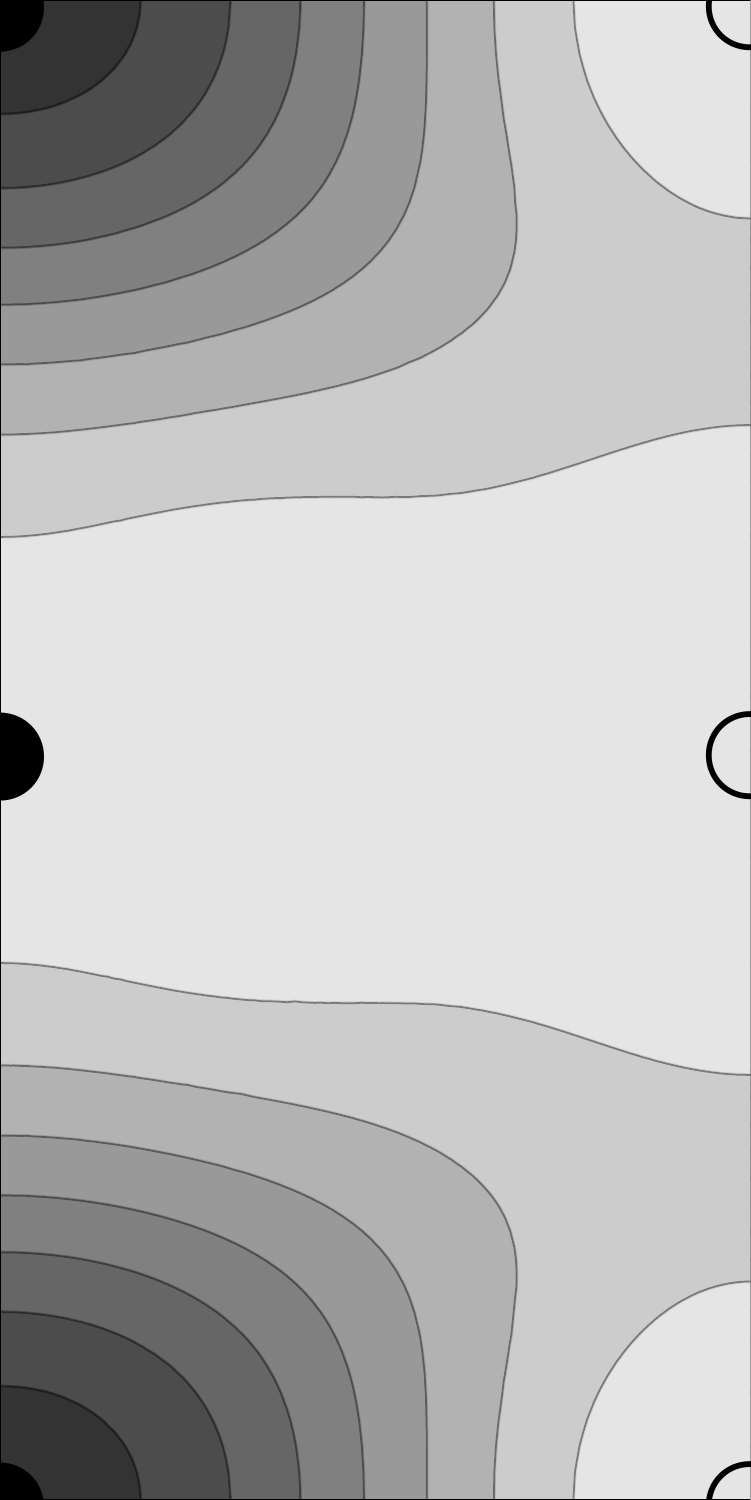} \hspace*{0.1cm} \includegraphics[height = .25 \textheight]{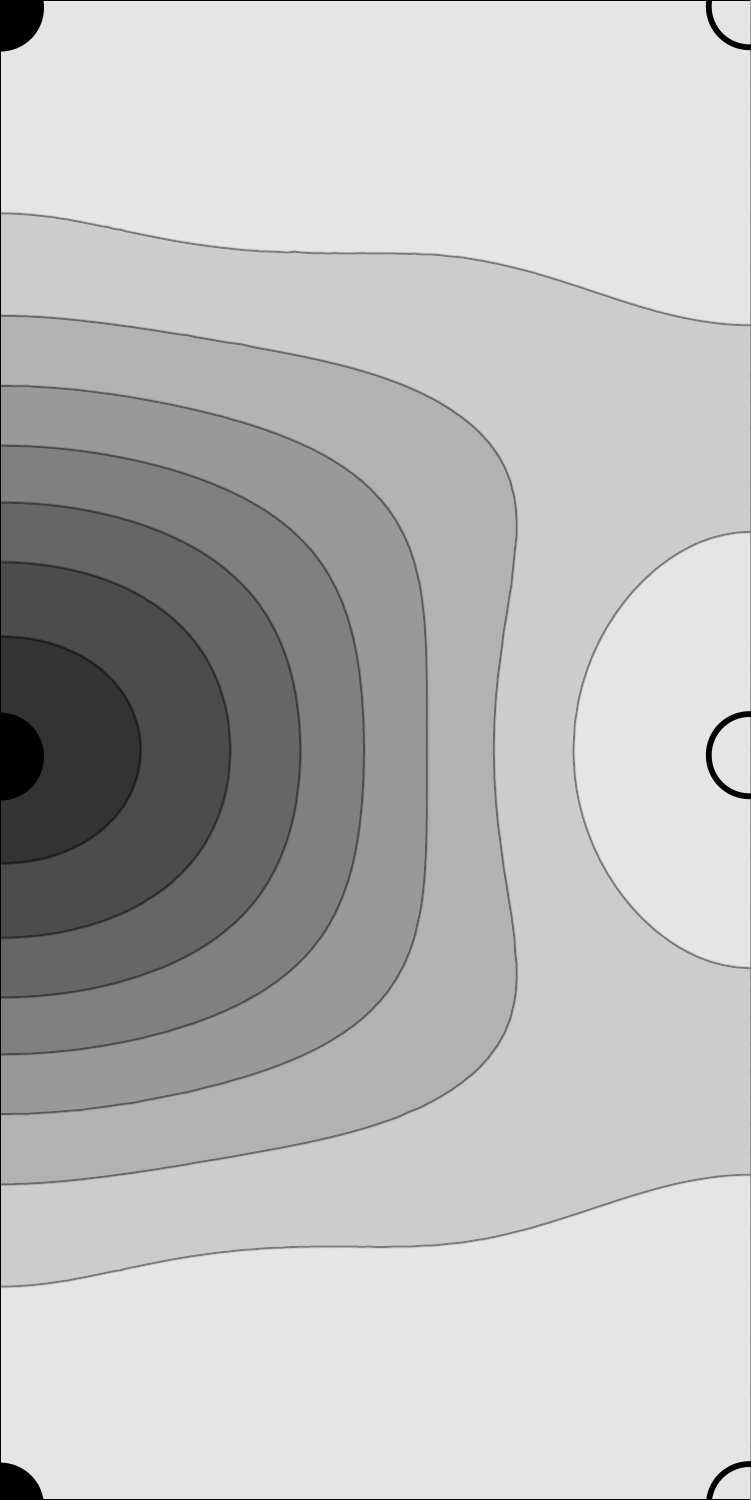} } \hspace*{1.0cm}
	\subfloat[\(k_1 = 0, k_2 = 1\)]{\includegraphics[height = .25 \textheight]{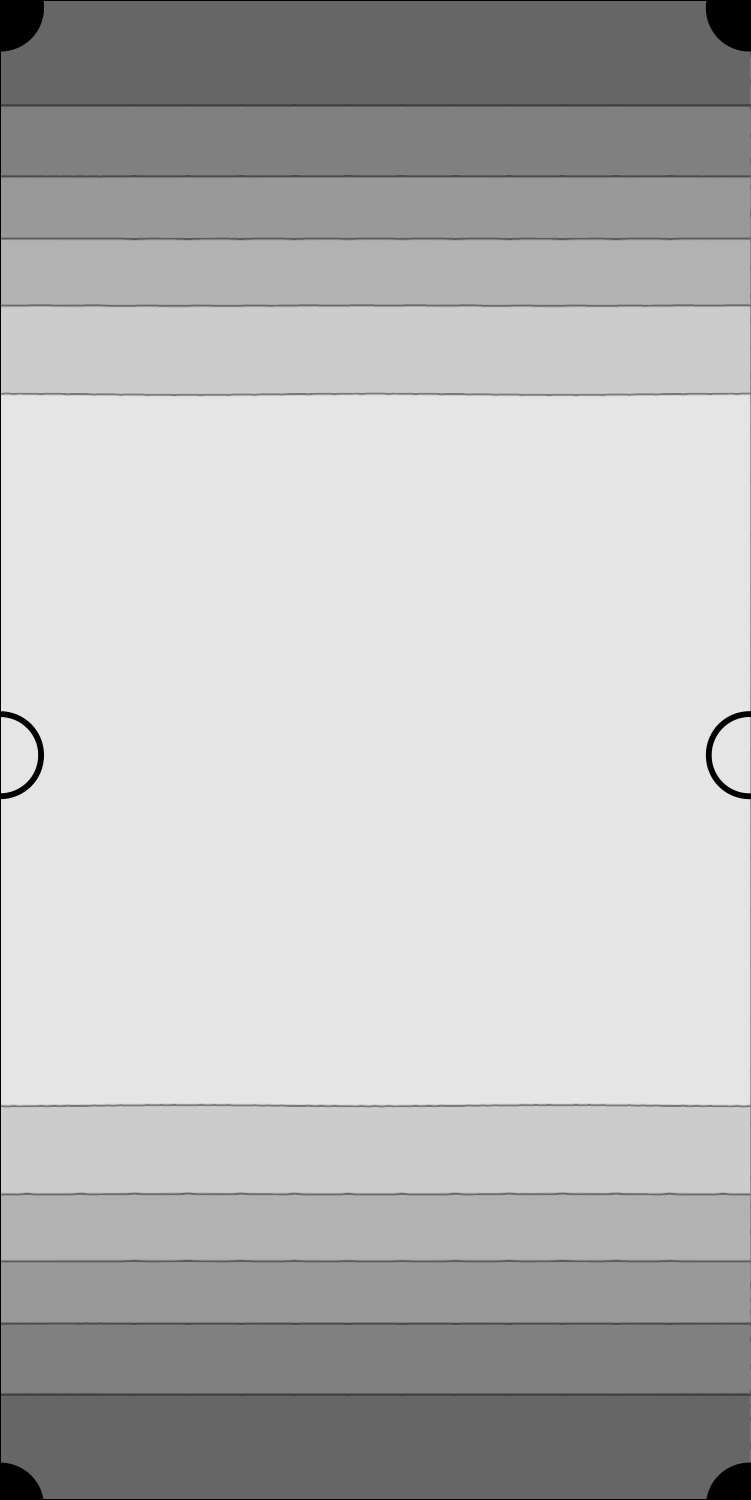} \hspace*{0.1cm} \includegraphics[height = .25 \textheight]{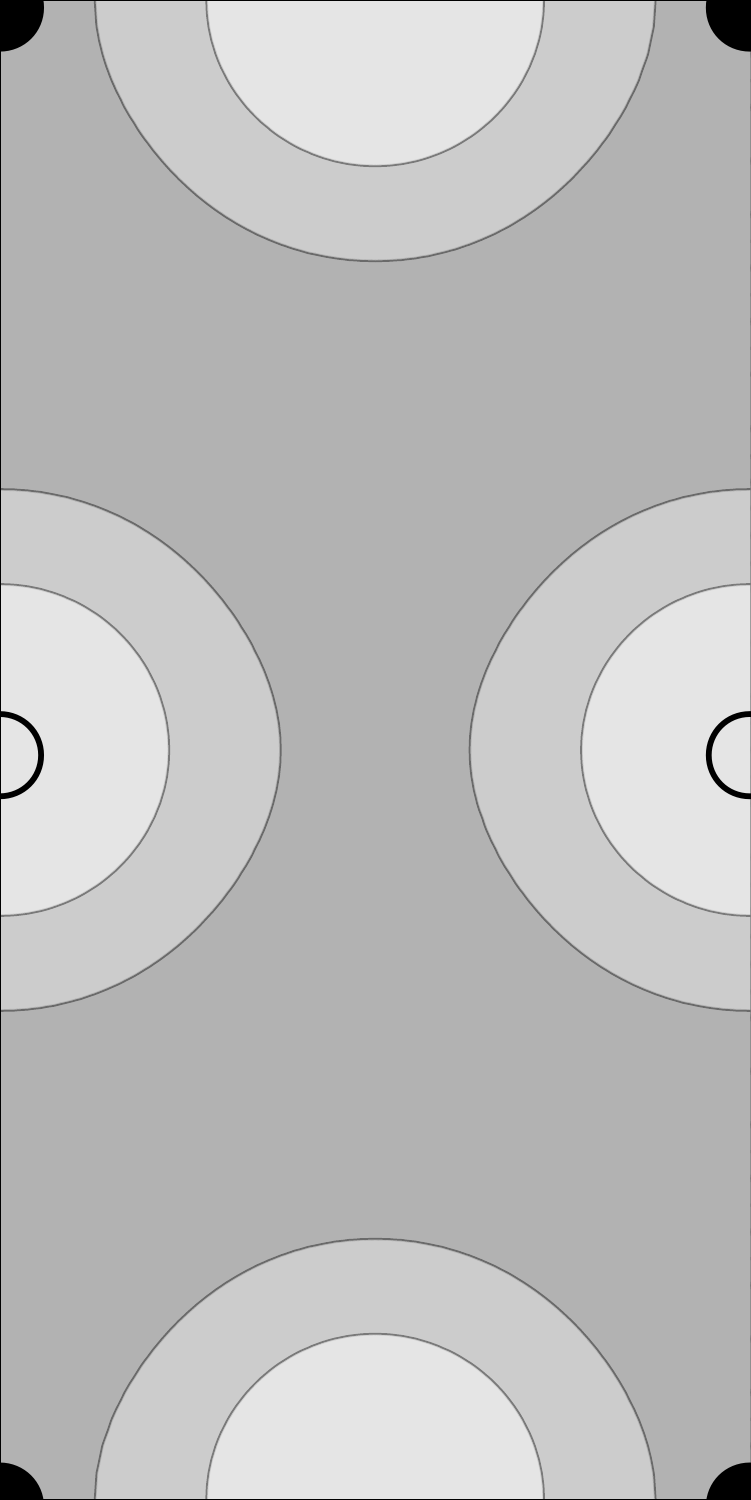} } \hspace*{0.1cm}
\\
	\subfloat[\(k_1 = 1, k_2 = 1\)]{\includegraphics[height = .25 \textheight]{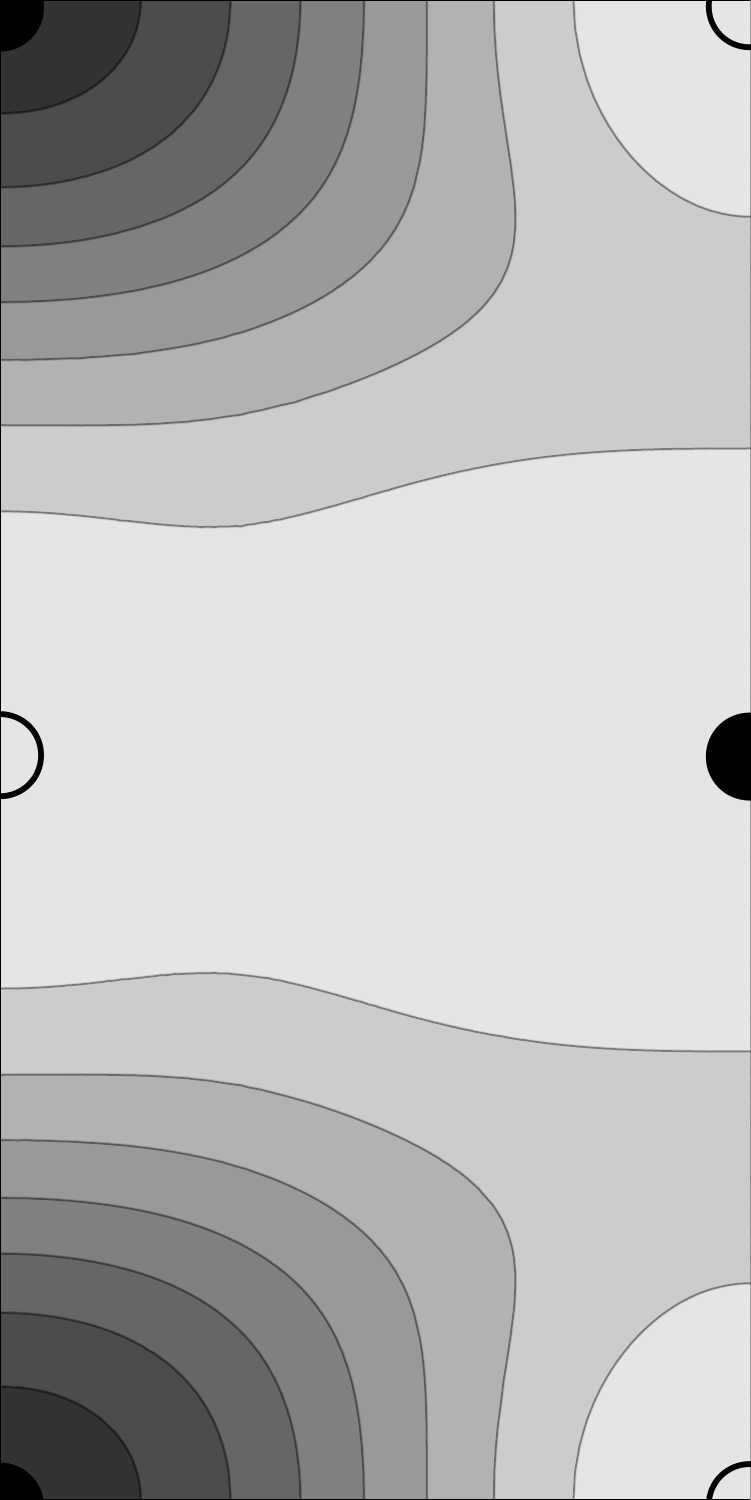} \hspace*{0.1cm} \includegraphics[height = .25 \textheight]{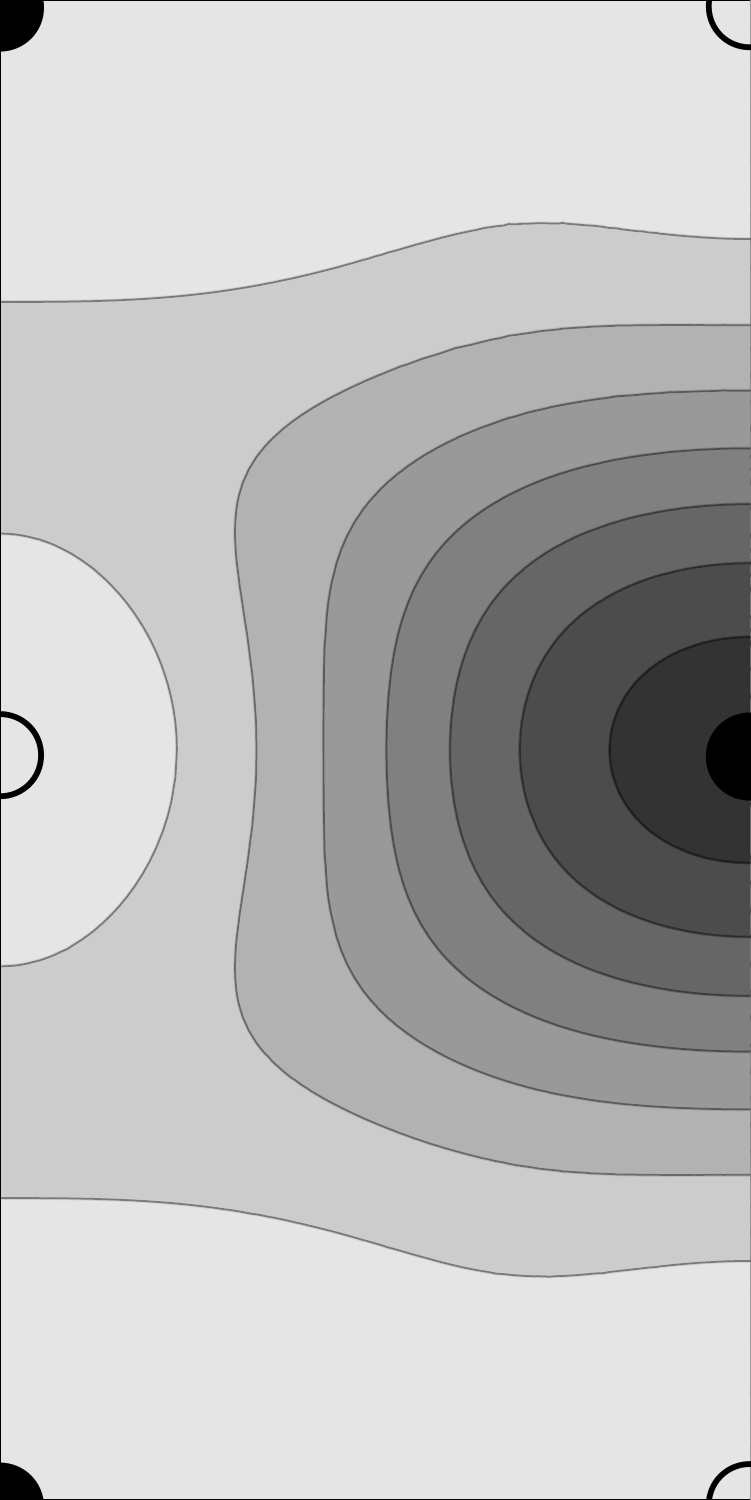} } \hspace*{0.1cm}
	\subfloat{\includegraphics[height = .25 \textheight]{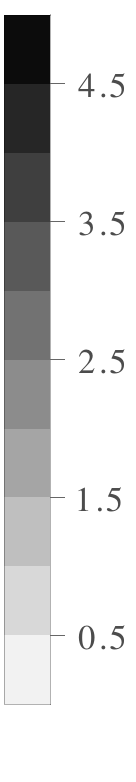}}
	\caption{The (absolute square of the) two possible wave functions for \(N = 2\) and non-trivial Wilson line configurations, classified by $k_m$ are depicted. Empty circles denote fixed points at which the wave function vanishes. \(j=0\,(1)\) is shown on the left(right).}
	\label{sec:wavefunctions:fig:wilson_lines}
	\end{center}
\end{figure}

Notably, not only the shape, but also the number of linear independent $P$-even and $P$-odd wave functions changes when Wilson lines are turned on. As anticipated, this reduces the number of independent zero modes from $(N+1)$ to $N$. Indeed, for non-zero Wilson lines, the range of \(j\) leading to linear independent, $P$-even wave functions is restricted to \(j \in \{ 0, \dots, N-1\}\). The same arguments show that the number of independent, $P$-odd wave functions is increased by one to a total of \(N\).

\section{Summary and Outlook}
\label{sec:Conclusion} 

We have studied supergravity in six dimensions compactified on the orbifold $T^2/\mathbb{Z}_2$. The standard discrete Wilson lines have been decomposed with respect to a canonical basis of 1-cycles which allow a direct interpretation in terms of singular, localized flux at the fixed points. The meaning of these ``canonical Wilson lines'' is independent of the presence of bulk flux and helps to understand  their physical effects. We expect that the use of canonical Wilson lines can also simplify the evaluation of anomalies in general gauge backgrounds as well as the treatment of discrete Wilson lines on other orbifolds.

In the case of chiral boundary conditions for 6d Weyl fermions and bulk flux, but without Wilson lines, we have considered the familiar 6d bulk and fixed point anomalies and we have computed the additional 4d chiral anomaly in the flux background. It is very satisfactory that the Green-Schwarz counter term designed to cancel the anomalies for vanishing flux gets modified such that it also cancels the additional contribution to the anomaly caused by the flux-induced zero modes.

The dimensional reduction to an effective four-dimensional theory reveals a particular form of the Green-Schwarz mechanism involving two axions. Depending on the background the two axions mix in a way that depends on the expectation values of the moduli fields. One linear combination, $\chi$, contributes to the vector boson mass via the Stueckelberg mechanism. The orthogonal combination, $a$, is massless and enters the Lagrangian through the standard coupling to the $U(1)$ field strength. Explicit expressions for $\chi$, $a$ and the vector boson mass were obtained as functions of the number of flux quanta and the expectation values of the moduli fields. It is interesting that the vector boson receives a mass even in the absence of an anomaly, due to a classical self-interaction in the flux background.

Finally, we have considered mass spectra and wave functions of charged bosons and fermions. They are fundamentally different in the cases with and without bulk flux. In particular, for non-zero flux, they do not depend on shape moduli and Wilson lines. We have constructed a convenient form of the fermionic zero mode wave functions on $T^2$ and $T^2/\mathbb{Z}_2$ in an arbitrary flux background. These wave functions and their multiplicity show a characteristic behavior if additional Wilson lines are switched on. The counting of the linear independent wave functions matches the expectations from orbifold projection and index theorem and their zeros are directly related to the localized flux. These results have important implications for model building. 

The present work suggests several further investigations. First, it will be interesting to extend our discussion of localized flux to general orbifolds $T^2/\mathbb{Z}_N$ where the canonical Wilson lines show further advantages. Second, one can extend our discussion to localized integer fluxes that will modify the spectrum of charged fields and their localization properties. Moreover, the evaluation of fixed point anomalies for more general internal spaces and gauge groups with arbitrary gauge backgrounds appears interesting. The investigation of the gravitational anomalies, in relation to the singular curvature at the fixed points, is another important aspect which we have not addressed in the present paper. Finally, a treatment of ``magnetized extra dimensions'' within a UV complete theory, in particular heterotic string theory appears very promising. These issues are currently under investigation \cite{bdrs1}.

Orbifolds with flux and Wilson lines contain all the ingredients needed to construct higher-dimensional models of grand unification: The flux can generate a multiplicity of generations, and symmetry breaking localized at the orbifold fixed points can break a grand unified group down to the Standard Model gauge group, thereby producing ``split multiplets'' that are needed for the Higgs sector. It is intriguing that such a pattern of symmetry breaking is also connected with the breaking of supersymmetry. These issues will be further discussed in a forthcoming paper \cite{bdrs2}.

\section*{Acknowledgments}
We thank Emilian Dudas, Stefan Groot Nibbelink, Jan Louis, Kai Schmidt-Hoberg, and especially
Arthur Hebecker for valuable discussions. This work was supported by the German Science Foundation (DFG) within the Collaborative Research Center (SFB) 676 ``Particles,
Strings and the Early Universe''. M.D. also acknowledges support from the Studienstiftung des deutschen Volkes.

}

\providecommand{\href}[2]{#2}\begingroup\raggedright\endgroup

\end{document}